\documentclass[%
 reprint,longbibliography,
 amsmath,amssymb,
 aps,superscriptaddress
]{revtex4-1}

\usepackage{graphicx}% Include figure files
\usepackage{dcolumn}% Align table columns on decimal point
\usepackage{bm}% bold math
\usepackage{hyperref}% add hypertext capabilities
\usepackage{xcolor}
\usepackage{amsmath, amssymb, amsfonts}
\usepackage{mathtools}
\usepackage{subfigure}
\usepackage{mathrsfs}
\newcommand{\col}[1]{\textcolor{red}{#1}}

\definecolor{purple1}{rgb}{128,0,128}

\newcommand{\bea}{\begin{eqnarray}}
\newcommand{\ea}{\end{eqnarray}}
\newcommand{\ord}{{\cal O}}

\begin{document}

\preprint{APS/123-QED}

\title{Stoner-Wohlfarth switching of the condensate magnetization in a dipolar spinor gas 
and the metrology of excitation damping} 
%Stoner-Wohlfarth switching in a dipolar spinor gas and the metrology of condensate 
%excitation damping}  
%How to Measure Dissipation Parameter $\Gamma$ for Spinor Bose-Einstein %Condensates Using Landau-Lifshitz-Gilbert Equation
% Force line breaks with \\

\author{Seong-Ho Shinn}

\affiliation{%
 Department of Physics and Astronomy, Seoul National University, 08826 Seoul, Korea}%

\author{Daniel Braun}

\affiliation{%
Eberhard-Karls-Universit\"at T\"ubingen, Institut f\"ur Theoretische Physik, 72076 T\"ubingen, Germany}

\author{Uwe R. Fischer}

\affiliation{%
 Department of Physics and Astronomy, Seoul National University, 08826 Seoul, Korea}
\date{\today}% It is always \today, today,
             %  but any date may be explicitly specified

\begin{abstract}
We consider quasi-one-dimensional  dipolar 
spinor Bose-Einstein condensates in the 
homogeneous-local-spin-orientation approximation, that is with unidirectional local magnetization. 
By analytically calculating the exact effective dipole-dipole interaction, we derive a Landau-Lifshitz-Gilbert equation for the dissipative condensate magnetization dynamics, and show how it leads to the Stoner-Wohlfarth model of a uni-axial ferro-magnetic particle, where 
the latter model determines the stable magnetization patterns and hysteresis curves for switching between them. 
For an external magnetic field pointing along the axial, long  direction, we analytically solve the Landau-Lifshitz-Gilbert equation. 
The solution explicitly demonstrates that the magnetic dipole-dipole interaction {\it accelerates} the dissipative dynamics of the  magnetic moment distribution and the associated dephasing of the magnetic moment direction. 
Under suitable conditions, 
dephasing of the magnetization direction due to dipole-dipole interactions 
occurs within time scales up to two orders of magnitude smaller than 
the lifetime of currently experimentally realized dipolar spinor condensates, e.g.,  produced with the large magnetic-dipole-moment atoms ${}^{166} \textrm{Er}$.
This enables experimental access to 
the dissipation parameter 
$\Gamma$ in the Gross-Pitaevski\v\i~mean-field equation, 
for a system currently lacking a  complete quantum kinetic treatment of dissipative processes  and, in particular,  an experimental check of the commonly used 
assumption that $\Gamma$ is a single scalar independent of spin indices.
\end{abstract}

\maketitle

\section{Introduction}
Ever since a phenomenological theory to describe the behavior of superfluid helium II near the $\lambda$ point has been developed by {Pitaevski\v\i} \cite{pitaevskii1959phenomenological}, 
% \qq in the intervening years %that's implied by the "ever since"
the dynamics of Bose-Einstein condensates (BEC) 
under dissipation has been intensely studied, see, e.g., \cite{WVLiu,Fedichev1998,PhysRevA.57.4057,Zaremba1999,Giorgini,Jackson2003,Proukakis2008}. 
Experimentally, the impact of Bose-Einstein condensation on excitation damping and its temperature dependence has for example been demonstrated in \cite{Jin1996,Jin1997,PhysRevLett.77.988,Stamper-Kurn1998}. 
%studied by various researchers. 
%From the data of \cite{PhysRevLett.77.988}, \cite{PhysRevA.57.4057} got $\Gamma \simeq %0.03$ for scalar BEC with sodium atoms where 

Dissipation in the form of condensate loss 
is defined by a dimensionless damping rate $\Gamma$ entering the left-hand side of the Gross-Pitaevski\v\i~equation, 
replacing the time derivative as %follows
$i\partial_t \rightarrow (i-\Gamma)\partial_t$. 
%is a dimensionless factor which %describes the degree of dissipation. 
While a microscopic theory of condensate damping is  
comparatively well established in the contact-interaction case, using various approaches,
cf., e.g., \cite{Zaremba1999,Nikuni2003,endo2011kinetic,PhysRevA.90.023631},    
%and to a relation of theoretically obtained damping rates of collective oscillations 
%to experiments is possible at least on the level of phenomenological and kinetic \cite{PhysRevA.57.4057,Khawaja}, 
we emphasize %the notable
the absence of a microscopic theory of damping in {\it dipolar spinor} gases. %\qq dipolar-spinor gases or dipolar spinor-gases ? one needs a hyphen and should decide which one. 
While for scalar dipolar condensates, partial answers as to the degree and origin of 
condensate-excitation damping %\qq hyphenized
have been found see, e.g., Refs.~\cite{Sykes,Natu,Wilson,Tercas}, in spinor or multicomponent gases the interplay of anisotropic long-range interactions and internal spinor or multicomponent degrees of freedom leads to a highly intricate and difficult-to-disentangle many-body behavior %\qq hyphenized
 of condensate-excitation damping. %\qq idem
 %of the distribution functions. 
% This is true in particular when a Bose-Einstein condensate is present, 
% that is in a quantum degenerate, highly nonclassical gas.}

In this paper, %by assuming that $\Gamma$ is independent of spin indices for spinor BEC, 
we propose a method to experimentally access 
$\Gamma$ in a dipolar spinor condensate 
by using the dynamics of the %condensate-density-scaled 
unidirectional local %(single domain) 
magnetization in a quasi-one-dimensional (quasi-1D) 
dipolar spinor BEC in the presence of an external magnetic field.  
To this end, we first derive an equation of motion for the magnetization of the BEC that has the form of a Landau-Lifshitz-Gilbert (LLG) equation \cite{Landau53,Gilbert2004,lakshmanan_fascinating_2011}, 
with an additional term due to the dipole-dipole interaction between the atoms.  
The LLG equation is ubiquitous in nano-magnetism, where it describes the creation and dynamics of %\qq magnetic domains
magnetization.  The static limit of this equation is,  
%leads to domains %that%we employ 
in the  limit of homogeneous local spin-orientation, 
described by the well-known Stoner-Wolfarth (SW) model \cite{stoner_mechanism_1948,SWreview,Hatomura} of a small magnetic particle with an easy axis of magnetization. We then investigate the magnetization switching after flipping the sign of the external magnetic field, and demonstrate the detailed dependence of  the switching dynamics 
on the dissipative parameter $\Gamma$.   

For a quasi-2D spinor BEC with inhomogeneous local magnetization, Ref.~\cite{PhysRevA.84.043607} has studied the magnetic domain wall formation 
process by deriving a LLG 
type equation. Here, we derive the  LLG equation in a quasi-1D spinor BEC with unidirectional 
local magnetization, in order to establish a most direct connection to the original 
SW model. 
In distinction to~\cite{Giovanazzi2004}, which studied the effective quasi-1D dipole-dipole interaction resulting from integrating out the two transverse directions 
within a simple approximation, 
 we employ below an exact analytic form of the dipole-dipole interaction.
 % without using any (uncontrolled) approximation.  
 %We show our derivation in Appendix~\ref{G_deriv}.
In Section~\ref{quasi_2d_GP_dissp_deriv}, we establish the quasi-1D spinor Gross-{Pitaevski\v\i} (GP) equation with dissipation, and equations of motion for the magnetization direction (unit vector) $\boldsymbol{M}$. 
Section~\ref{connec_to_SW} %studies the dynamics of $\boldsymbol{M}$ within the
shows how the LLG equation and the 
SW model result, 
and  Section~\ref{numerical_res} derives 
{%\it
  analytical} solutions to the equations of motion for $\boldsymbol{M}$ 
when the external magnetic field points along the long, $z$ axis.
%, hence we suggest a method to measure $\Gamma$ in experiment.
We summarize our results in section~\ref{concl}. 

We defer two longer derivations to Appendices.
The analytical form of the 
effective dipole-dipole interaction energy 
%for %the quasi-1D dipolar %\qq gas %\qq
is deduced 
in Appendix~\ref{G_deriv}, and the quasi-1D GP 
%Gross-Pitaevski\v\i~
mean-field 
equation with dissipation is described in detail in Appendix~\ref{quasi_1d_llg_deriv}.  
Finally, in  Appendix~\ref{Gamma_tensor}, we briefly discuss to which extent
relaxing the usual 
simplifying assumption that dissipation even in the spinor case is described 
by a single scalar changes the LLG equation, and whether this affects the SW model and its predictions.
% regarding the relaxation dynamics of magnetization.

\section{\label{quasi_2d_GP_dissp_deriv}General description of damping in BECs}
%Relation of damping phenomenology to microscopics}
The standard %reference on
derivation of 
the quantum kinetics of Bose-Einstein condensate damping 
\cite{Zaremba1999} starts from the microscopic Heisenberg equation of motion for the quantum field operator $\hat{\psi} \left( \boldsymbol{r}, t \right)$, for a scalar (single component) 
BEC in the $s$-wave scattering limit.  
Using their results, \cite{PhysRevA.67.033610} obtained a 
mean-field equation to describe the dissipation of scalar BEC, 
whose form is
\begin{equation}\label{eq:1}
\left( i - \Gamma \right) \hbar \frac{\partial \psi}{\partial t} = H \psi  
\end{equation}
 where $\psi$ is the (in the large $N$ limit)  dominant 
 mean-field part upon expanding the full bosonic field operator $\hat{\psi}$. 

In Ref.~\cite{pitaevskii1959phenomenological}, Pitaevski\v\i~ 
obtained a similar but slightly different form of the dissipative mean-field equation 
based on phenomenological considerations, $i \hbar \frac{\partial \psi}{\partial t} = \left(1 - i \Gamma \right) H \psi$, by %assuming that
parametrizing the deviation from exact continuity for the condensate fraction while minimizing the energy~\cite{pitaevskii1959phenomenological}.  The latter deviation %from equilibrium 
is assumed to be small, which is equivalent to assuming that $\Gamma$ remains small. 
This provides a clear physical interpretation %\qq to
of the damping mechanism, namely one based on particle loss from the condensate fraction. 
The version of Pitaevski{\v\i}
can be written as
  \begin{equation}
    \label{eq:2}
\left( i - \Gamma \right) \hbar \frac{\partial \psi}{\partial t} = \left( 1 + \Gamma^2 \right) H \psi\,.
  \end{equation} 
It can thus be simply obtained by rescaling time with a factor $1+\Gamma^2$ compared to 
\eqref{eq:1}. 
%Zaremba et al.'s version. 
Hence, as long as one does not predict precisely $\Gamma$, the two 
dissipative equations \eqref{eq:1} and \eqref{eq:2} cannot be distinguished experimentally from the dynamics they induce. 
From the data of \cite{PhysRevLett.77.988}, \cite{PhysRevA.57.4057}
estimated typical values of $\Gamma \simeq 0.03$ for a {\it scalar} BEC of
${}^{23} \textrm{Na}$ 
atoms (see also \cite{Stamper-Kurn1998}),  
%Thus, $\left( 1 + \Gamma^2 \right)$ 
%for $\Gamma \simeq 0.03$, this factor 
%is close to $1$, and there is little hope to distinguish between
which shows that to distinguish between \eqref{eq:1} and \eqref{eq:2} experimentally 
the theoretical predictions of $\Gamma$ would need to be precise to the order of $10^{-4}$. %is a difficult task.   

How eqs.\eqref{eq:1} and \eqref{eq:2} can be generalized to the 
dipolar spinor gases is comparatively little investigated. 
%According to~\cite{Zaremba1999} and~\cite{PhysRevA.67.033610}, $\Gamma$ derives from the three-field correlation function $\langle \delta \hat{\psi}^{\dagger} \left( \boldsymbol{r}, t \right) \delta \hat{\psi} \left( \boldsymbol{r}, t \right) \delta \hat{\psi} \left( \boldsymbol{r}, t \right) \rangle $
%assuming to use some
%in a basis %\qq (not mentioned specifically though) such that
%where $\langle \delta \hat{\psi} \left( \boldsymbol{r}, t \right) \delta \hat{\psi} \left( \boldsymbol{r}, t \right) \rangle = 0$ 
%where $\hat{\psi} \left( \boldsymbol{r}, t \right) = \psi \left( \boldsymbol{r}, t \right) + \delta \hat{\psi} \left( \boldsymbol{r}, t \right)$, with % and $\psi \left( \boldsymbol{r}, t \right)$ is
%expectation value of $\hat{\psi} \left( \boldsymbol{r}, t \right)$. i.e., 
%$ \psi \left( \boldsymbol{r}, t \right) = \langle \hat{\psi} \left( \boldsymbol{r}, t \right) \rangle $ and $\langle \delta \hat{\psi} \left( \boldsymbol{r}, t \right) \rangle= 0$ in the symmetry-breaking type of mean-field approach. 
Using a symmetry-breaking mean-field approach by writing the quantum field operator 
as $\hat{\psi} \left( \boldsymbol{r}, t \right)$ as 
$\hat{\psi} \left( \boldsymbol{r}, t \right) = \psi \left( \boldsymbol{r}, t \right) + \delta \hat{\psi} \left( \boldsymbol{r}, t \right)$, with $\psi \left( \boldsymbol{r}, t \right) = \langle \hat{\psi} \left( \boldsymbol{r}, t \right) \rangle $ and $\langle \delta \hat{\psi} \left( \boldsymbol{r}, t \right) \rangle= 0$, 
\cite{Zaremba1999} and~\cite{PhysRevA.67.033610} showed that $\Gamma$ is derived 
from the three-field correlation function $\langle \delta \hat{\psi}^{\dagger} \left( \boldsymbol{r}, t \right) \delta \hat{\psi} \left( \boldsymbol{r}, t \right) \delta \hat{\psi} \left( \boldsymbol{r}, t \right) \rangle $ in a basis where $\langle \delta \hat{\psi} \left( \boldsymbol{r}, t \right) \delta \hat{\psi} \left( \boldsymbol{r}, t \right) \rangle = 0$.
From this microscopic origin, based on correlation functions, it is clear that in principle
$\Gamma$ might depend on the spin indices in a spinor BECs and hence become
a tensor (see Appendix \ref{Gamma_tensor} for a corresponding
phenomenological generalization).
Nevertheless, it is commonly assumed cf.,e.g.,~\cite{JiHQV,PhysRevA.84.043607}, 
that $\Gamma$ does not depend on spin indices, and the scalar value
found specifically in  \cite{PhysRevA.57.4057} for a {\it scalar} BEC of  ${}^{23}$ Na
atoms is commonly used, %\qq and
while a clear justification of this assumption is missing. 

Extending the microscopic derivations in \cite{Zaremba1999} and~\cite{PhysRevA.67.033610}
to the spinor case would be theoretically interesting, but is %rather well 
beyond the scope  of the present paper.  Here, we instead focus on the question
whether the standard assumption that the damping of each spinor
component can be described by the mean-field equation
\cite{PhysRevA.67.033610} leads to experimentally falsifiable
dynamical signatures.  It will turn out that this assumption
introduces an additional strong dephasing in the spin-degrees
of freedom, amplified by the dipolar interaction.  Hence, even on
time scales on which the decay of the condensate fraction according to
\eqref{eq:1} can be neglected, the relaxation of the magnetization of
the BEC potentially offers valuable insights whether the scalar-$\Gamma$ assumption is
justified. 
Indeed, in \cite{Er_quasi_1D_exp} it was shown experimentally that on
the time scale of the switching dynamics of the magnetization the number of particles in
the condensates remains approximately constant. One might wonder, then, 
which dissipative mechanism is left.  However, as we will show, 
by assuming the same GP 
%Gross-Pitaevsk\v\i~
equation for each component
of the spinor as for scalar bosons, additional dephasing occurs that
is in fact much more rapid than the decay of condensate density due to
% an accelerated
dephasing accelerated by the dipole-dipole interaction.

%\section{
%\label{quasi_2d_GP_dissp_deriv}Quasi-one-dimensional dissipative Gross-Pitaevskii Equation}

\section{Mean-field dynamics of damping in dipolar spinor BECs}
For a spinor BEC, linear and quadratic Zeeman interactions are commonly included in the Hamiltonian. %Among these, t
The quadratic Zeeman interaction is related to a second-order perturbation term in the total energy that can be induced by the interaction with an external magnetic field ($q_{B}$) as well as with the interaction with a microwave field ($q_{\rm MW}$)~\cite{Uedareview}. Specifically, by applying a linearly polarized microwave field, one can change $q_{\rm MW}$ without changing $q_{B}$~\cite{PhysRevA.73.041602,PhysRevA.79.043631}. Hence, we will assume that the quadratic Zeeman term can be rendered zero by suitably changing $q_{\rm MW}$.

Following \cite{PhysRevA.84.043607}, we thus assert that for a dipolar spinor
BEC without quadratic Zeeman term, the mean-field equation can
be written as 
\begin{widetext}
\begin{eqnarray}
\left( i - \Gamma \right) \hbar \frac{\partial \psi \left( \boldsymbol{r}, t \right)}{\partial t} = && \; 
\left\lbrack 
- \frac{\hbar^2}{2 m} \nabla^2 
+ V_{\rm tr} \left( \boldsymbol{r} \right) 
+ c_0 \left\vert \psi \left( \boldsymbol{r}, t \right) \right\vert^2 
- \hbar 
\left\{
\boldsymbol{b} 
- \boldsymbol{b}_{dd} \left( \boldsymbol{r}, t \right) 
\right\} 
\cdot 
\boldsymbol{\hat{f}} 
\right\rbrack 
\psi \left( \boldsymbol{r}, t \right)
\nonumber\\
&& \quad + 
\sum_{k = 1}^{S} c_{2 k} 
\sum_{\nu_1, \nu_2, \cdots, \nu_k = x, y, z} 
F_{\nu_1, \nu_2, \cdots, \nu_k} \left( \boldsymbol{r}, t \right) 
\hat{f}_{\nu_1} \hat{f}_{\nu_2} \cdots \hat{f}_{\nu_k} 
\psi \left( \boldsymbol{r}, t \right).
\label{GP_SW_q}
\end{eqnarray}
\end{widetext}
where $\psi \left( \boldsymbol{r}, t \right)$ is a vector quantity whose $\alpha$-th component in the spinor basis is $\psi_{\alpha} \left( \boldsymbol{r}, t \right)$ (spin-space indices from the beginning of the Greek alphabet such as $\alpha,\beta,\gamma,\ldots$ are integers running from $-S$ to $S$). 
In this expression, $\hbar \boldsymbol{\hat{f}}$ is the spin-$S$ operator where the spin ladder is defined %\qq from
by $\hat{f}_{z} \left\vert \alpha \right\rangle = \alpha \left\vert \alpha \right\rangle$ and $\left\langle \alpha \vert \beta \right\rangle = \delta_{\alpha, \beta}$,  while 
$F_{\nu_1, \nu_2, \cdots, \nu_k} \left( \boldsymbol{r}, t \right) \coloneqq 
\psi^{\dagger} \left( \boldsymbol{r}, t \right) 
\hat{f}_{\nu_1} \hat{f}_{\nu_2} \cdots \hat{f}_{\nu_k} 
\psi \left( \boldsymbol{r}, t \right)$ 
are the components of the expectation value of 
$\hat{f}_{\nu_1} \hat{f}_{\nu_2} \cdots \hat{f}_{\nu_k}$. 
%$k$-th nematic tensor,  %{density}, 
The Larmor frequency vector reads 
$\boldsymbol{b} = g_F \mu_B \boldsymbol{B} / \hbar$ (with Land\'e
g-factor $g_F$, Bohr magneton $\mu_B$, and the external magnetic 
%field
induction %is
$\boldsymbol{B}$), 
%and 
$\hbar\boldsymbol{b}_{dd} \left( \boldsymbol{r}, t \right) \cdot \boldsymbol{e}_{\nu} = 
c_{dd} 
%\int d \boldsymbol{r'} 
\int d^3 r' \; 
\sum_{\nu' = x, y, z} 
Q_{\nu, \nu'} \left( \boldsymbol{r} - \boldsymbol{r'} \right) 
F_{\nu'} \left( \boldsymbol{r'}, t \right).$ 
Here, 
$c_{dd} = \mu_{0} \left( g_{F} \mu_{B} \right)^2 / \left( 4 \pi \right)$ and 
$\boldsymbol{e}_{\nu}$ is a unit vector along the $\nu$ axis ~\cite{Uedareview}
 (by convention, indices from the middle of the Greek alphabet 
such as $\kappa,\lambda,\mu,\nu,\ldots = x,y,z$ denote spatial indices), and $Q_{\nu, \nu'}$ is the spin-space tensor 
defined in Eq.~\eqref{def_Q_nu_nu} of Appendix \ref{G_deriv}. 
Finally, $m$ is the boson mass, $c_0$ %\qq is
the density-density interaction coefficient, 
and $c_{2k}$ %\qq is 
%{$\hat{f}_{\nu_1} \hat{f}_{\nu_2} \cdots \hat{f}_{\nu_k}$} 
%$k$-th nematic tensor 
the interaction coefficient parametrizing the spin-spin interactions, where $k$ is an positive integer 
running from 1 to $S$ \cite{PhysRevA.84.043607}. 
For example, $c_{2}$ is the spin-spin interaction coefficient of a spin-1 gas ($S=1$).  

\begin{figure} [b]
\centering
\includegraphics[width=0.48\textwidth]{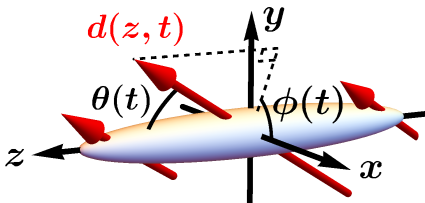}
\caption{Schematic of the considered geometry in a quasi-1D gas (shaded ellipsoid). 
The length of the red magnetization arrows, all pointing in the same direction 
(homogeneous local-spin-orientation 
limit),  represents $\left\vert \boldsymbol{d} \left( z, t \right) \right\vert$.}
% at its $z$ axis location. 
%As this figure shows, the magnitude of local magnetic dipole moment $\boldsymbol{d} \left( z, t %\right)$ may change depending on the position $z$, but its direction is same for every $z$.
\label{d-singledomain-fig}
\end{figure}

To develop a simple and intuitive physical approach, we consider a quasi-1D gas 
for which one can perform analytical calculations. 
We set the trap potential as 
%We consider a quasi-1D gas (with long axis directed along the $z$ axis and strongly confined perpendicularly), 
%with trap potential given by 
\begin{eqnarray}
V_{\rm tr} \left( x, y, z \right) = \frac{1}{2} m \omega_{\perp}^2 \left( x^2 + y^2 \right) + V \left( z \right)\,,
\label{trap}
\end{eqnarray}
so that the long axis of our gas is directed along the $z$ axis and the gas is strongly confined perpendicularly. 

For a harmonic trap along all directions, i.e. when $V(z)= m \omega_z^2 z^2 / 2$, we set $\omega_\perp \gg \omega_z$. 
For a box trap along $z$, i.e. when $V \left( z \right) = 0$ for $\left\vert z \right\vert \le L_z$ and $V \left( z \right) = \infty$ for $\left\vert z \right\vert > L_z$, our gas will be strongly confined 
along $z$ as long as 
%tje 
the quasi-1D condition is satisfied, where we will discuss below whether the  condition is satisfied, in section~\ref{Box-trap-consideration}. 
%$\omega_\perp \gg \omega_z$ for harmonic trapping along all directions, i.e. when $V(z)=\frac12 \omega_z z^2$. 

Single-domain spinor BECs have been already realized, for example, using spin-1 ${}^{87} \textrm{Rb}$~\cite{Palacios_2018}. 
This single-domain approximation is %\qq rather
  common in nanomagnetism, see for example \cite{SWreview}, by 
assuming magnetic particles much smaller than the typical width of a domain wall.  
The local magnetization is related to the expectation value 
$\hbar \boldsymbol{F} \left( \boldsymbol{r}, t \right)\equiv \hbar\psi^\dagger(\bm r,t) \hat{\bm f}\psi(\bm r,t)$ of the 
spatial spin density operator 
by $\boldsymbol{d} \left( \boldsymbol{r}, t \right) = g_{F} \mu_{B} \boldsymbol{F} \left( \boldsymbol{r}, t \right)$.
An unidirectional local magnetization $\boldsymbol{d} \left( z, t \right)$ is then given by  
\begin{eqnarray}
d_x \left( z, t \right) 
= && \; d \left( z, t \right) 
\sin \theta \left( t \right) \cos \phi \left( t \right), 
\nonumber\\
d_y \left( z, t \right) 
= && \; d \left( z, t \right) 
\sin \theta \left( t \right) \sin \phi \left( t \right), 
\label{d_def} \\
d_z \left( z, t \right) 
= && \; d \left( z, t \right) 
\cos \theta \left( t \right), 
\nonumber
\end{eqnarray} 
where 
$d_{\nu} \left( z, t \right) = \boldsymbol{d} \left( z, t \right) \cdot \boldsymbol{e}_{\nu}$ is the 
$\nu$-th component of $\boldsymbol{d} \left( z, t \right)$, 
$d \left( z, t \right) = \left\vert \boldsymbol{d} \left( z, t \right) \right\vert$, 
$\theta \left( t \right)$ is polar angle of $\boldsymbol{d} \left( z, t \right)$, and 
$\phi \left( t \right)$ is azimuthal angle of $\boldsymbol{d} \left( z, t \right)$. 
For an illustration of the geometry considered, see  
Fig.~\ref{d-singledomain-fig}. 
For a single component dipolar BEC, $\boldsymbol{F} \left( \boldsymbol{r}, t \right)$ has a fixed direction. To study the relation of the Stoner-Wohlfarth model, in which $\boldsymbol{F} \left( \boldsymbol{r}, t \right)$ changes its direction, with a dipolar BEC, a 
multi-component dipolar BEC should therefore be employed.

%so that the system is quasi-1D in $z$ direction, 
%Motivated by~\cite{GPfortran}, 
%this is a physical assumption.
%Are you saying it is motivated by the availability of a software
%package?
%This ansatz is motivated by 
%commonly used 
%the approach in \cite{GPfortran} for a quasi-1D system without dissipation, and simplifies . 
%Though it is about numerical calculation, their ansatz simplifies the 
%the analytic form of the mean-field equation. 

%Let $\psi_\alpha \left( \boldsymbol{r}, t \right)$ be the $\alpha$-th component of the mean-field wavefunction $\psi \left( \boldsymbol{r}, t \right)$ (where spin-space indices from the beginning of the Greek alphabet such as $\alpha,\beta,\gamma,\ldots$ are integers running from $-S$ to $S$). Then, one 

In the quasi-1D approximation, the order parameter $\psi_{\alpha} \left( \boldsymbol{r}, t \right)$ is commonly assumed to be of the form 
\begin{eqnarray}
\psi_{\alpha} \left( \boldsymbol{r}, t \right) = \frac{e^{- \rho^2 / \left( 2 l_{\perp}^2 \right)}}{l_{\perp} \sqrt{\pi}} 
\Psi_{\alpha} \left( z, t \right) .
\end{eqnarray}
where $l_{\perp}$ is the harmonic oscillator length in the $x--y$ plane and $\rho = \sqrt{x^2 + y^2}$. 
Assuming our gas is in the 
homogeneous local spin-orientation 
limit, we may also apply a 
single mode approximation in space so that $\Psi_{\alpha} \left( z, t \right) = \Psi_{\rm uni} \left( z, t \right) \zeta_{\alpha} \left( t \right)$. 
The time-dependent spinor part is 
\begin{eqnarray}
\zeta_{\alpha} \left( t \right) 
= \left\langle \alpha \right\vert e^{- i \hat{f}_{z} \phi \left( t \right)} e^{- i \hat{f}_{y} \theta \left( t \right)} 
\left\vert S \right\rangle,
\label{wf_zeta_def}
\end{eqnarray} 
for spin-$S$ particles~\cite{Uedareview,PhysRevA.84.043607} and the normalization reads 
$\left\vert \zeta \left( t \right) \right\vert^2 \coloneqq\zeta^{\dagger} \left( t \right) \zeta \left( t \right) = 1$. 
Finally, due to the $\left( i - \Gamma \right)$ factor on the left-hand side of Eq.~\eqref{GP_SW_q}, for the ease of calculation, we may make the following ansatz for the $\psi_{\alpha} \left( \boldsymbol{r}, t \right)$, cf.~Ref.~\cite{GPfortran}, 
%may make the following ansatz for the $\psi_{\alpha} \left( \boldsymbol{r}, t \right)$, cf.~Ref.~\cite{GPfortran} 
\begin{eqnarray}
\psi_{\alpha} \left( \boldsymbol{r}, t \right) = && \; 
\frac{e^{- \rho^2 / \left( 2 l_{\perp}^2 \right)}}{l_{\perp} \sqrt{\pi}} \Psi \left( z, t \right) \zeta_{\alpha} \left( t \right) e^{- \left( i + \Gamma \right) \omega_{\perp} t / \left( 1 + \Gamma^2 \right)} . 
\nonumber\\
\label{quasi_1d_wavefunc}
\end{eqnarray}
%Specifically, if we denote $\psi_\alpha \left( \boldsymbol{r}, t \right)$ as the $\alpha$-th component of $\psi \left( \boldsymbol{r}, t \right)$, 
%\begin{eqnarray}
%\psi_{\alpha} \left( \boldsymbol{r}, t \right) = && \; 
%\frac{e^{- \rho^2 / \left( 2 l_{\perp}^2 \right)}}{l_{\perp} \sqrt{\pi}} \Psi \left( z, t \right) \zeta_{\alpha} \left( t \right) e^{- \left( i + \Gamma \right) \omega_{\perp} t / \left( 1 + \Gamma^2 \right)}, 
%\nonumber\\
%\label{quasi_1d_wavefunc-psim}
%\end{eqnarray}
%where $\zeta_\alpha \left( t \right) = \left\langle \alpha \right\vert e^{- i \hat{f}_{z} \phi \left( t \right)} e^{- i \hat{f}_{y} \theta \left( t \right)} \left\vert S \right\rangle$.
From our ans\"atze in Eq.~\eqref{wf_zeta_def} %\qq how is that one an ansatz?
and~\eqref{quasi_1d_wavefunc}, one concludes that the %mean-field 
expectation value of the (spatial) spin-density operator is 
%\begin{widetext} 
\begin{eqnarray}
\hbar F_x \left( \boldsymbol{r}, t \right) 
= && \; \hbar S \frac{e^{- \rho^2 / l_{\perp}^2}}{\pi l_{\perp}^2} \left\vert \Psi \left( z, t \right) \right\vert^2 e^{- 2 \Gamma \omega_{\perp} t / \left( 1 + \Gamma^2 \right)} 
\nonumber\\& & \times 
\sin \theta \left( t \right) \cos \phi \left( t \right), 
\nonumber\\
\hbar F_y \left( \boldsymbol{r}, t \right) 
= && \; \hbar S \frac{e^{- \rho^2 / l_{\perp}^2}}{\pi l_{\perp}^2} \left\vert \Psi \left( z, t \right) \right\vert^2 e^{- 2 \Gamma \omega_{\perp} t / \left( 1 + \Gamma^2 \right)} 
\nonumber\\& & \times
\sin \theta \left( t \right) \sin \phi \left( t \right), 
\nonumber\\
\hbar F_z \left( \boldsymbol{r}, t \right) 
= && \; \hbar S \frac{e^{- \rho^2 / l_{\perp}^2}}{\pi l_{\perp}^2} \left\vert \Psi \left( z, t \right) \right\vert^2 e^{- 2 \Gamma \omega_{\perp} t / \left( 1 + \Gamma^2 \right)} 
\nonumber\\& & \times
\cos \theta \left( t \right). 
%\nonumber 
\label{F_expect_val}
\end{eqnarray}
%\end{widetext} 
The above equations lead to unidirectional local magnetization,  
which has been assumed in Eqs.~\eqref{d_def}, in the quasi-1D limit (after integrating out  the strongly confining $x$ and $y$ axes). 
%Unidirectional local magnetization corresponds to the single-domain approximation familiar in nanomagnetism \cite{SWreview}. 
Note however that our ansatz in Eq.~\eqref{quasi_1d_wavefunc} is sufficient, but not necessary 
for the 
homogeneous-local-spin-orientation 
%single-domain 
limit, and the 
homogeneous-local-spin-orientation 
ansatz is thus designed to render our approach as simple as possible. 

Because we are not assuming any specific form of $\Psi \left( z, t \right)$ in our ansatz in Eq.~\eqref{quasi_1d_wavefunc},%\qq  so that  %Therefore, our ansatz 
we cover every possible time behavior of $\left\vert \psi \left( \boldsymbol{r}, t \right) \right\vert^2
 \coloneqq\psi^{\dagger} \left( t \right) \psi \left( t \right) $: 
\begin{eqnarray}
\left\vert \psi \left( \boldsymbol{r}, t \right) \right\vert^2 
= && \; \frac{e^{- \rho^2 / l_{\perp}^2}}{\pi l_{\perp}^2} \left\vert \Psi \left( z, t \right) \right\vert^2 e^{- 2 \Gamma \omega_{\perp} t / \left( 1 + \Gamma^2 \right)} .
\label{psi_sq_explicit}
\end{eqnarray}
Eq.~\eqref{psi_sq_explicit} explicitly shows that 
Eq.~\eqref{quasi_1d_wavefunc} does not imply an
exponentially decaying wavefunction with time 
since $\left\vert \Psi \left( z, t \right) \right\vert^2$ can be any physical function of time $t$. 
However, the ansatz \eqref{quasi_1d_wavefunc} simplifies the resulting equation for $\Psi(z,t)$, Eq.\eqref{quasi_1d_gp_diss_maintext} below.    

\begin{figure} [t]
\centering
\subfigure{%\label{phi_p_phase}
\includegraphics[width=0.4\textwidth]{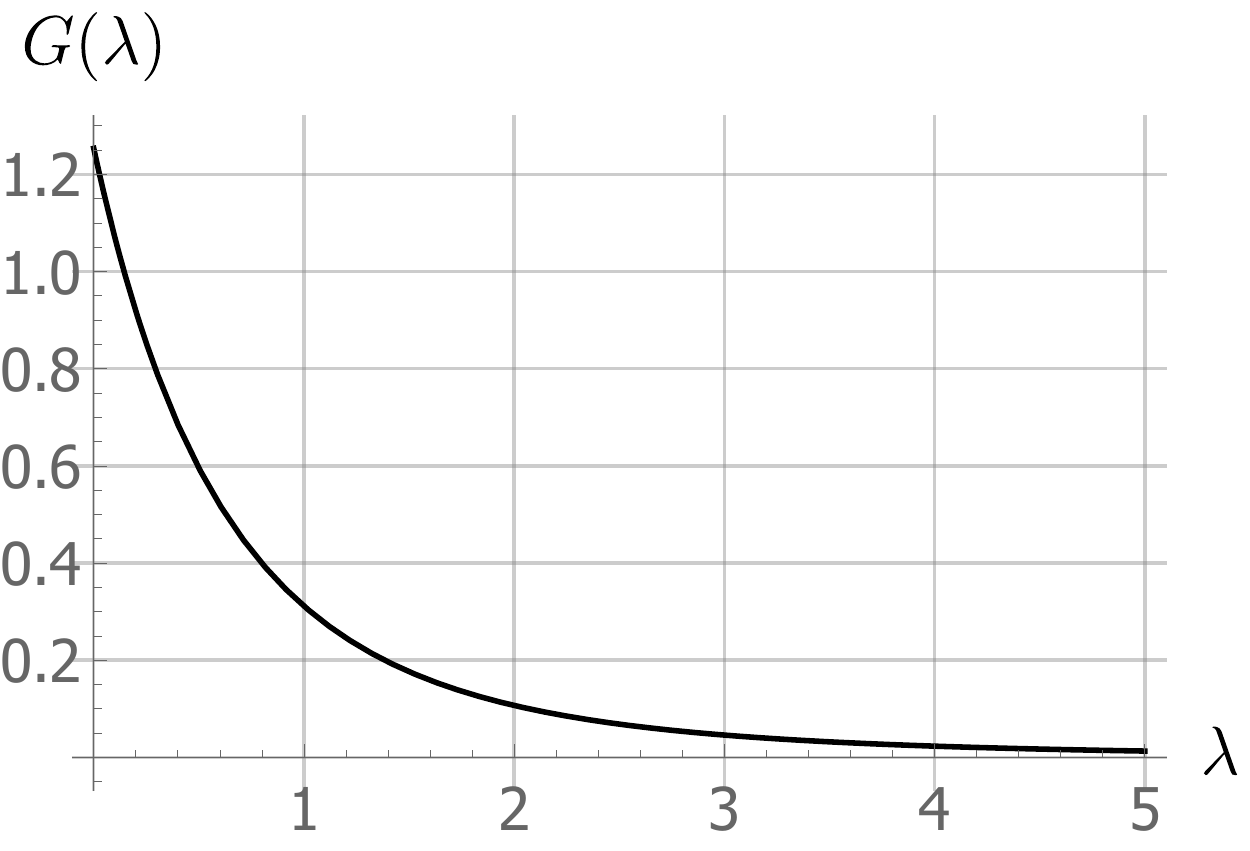}}
\caption{\label{G-plot} The function $G \left( \lambda \right)$ 
defined in Eq.~\eqref{F_defin}.  Note that $G \left( \lambda \right) \simeq 2 / \lambda^3 + \ord\left( \lambda^{-5} \right)$ for %$\lambda \rightarrow \infty$, 
$\lambda \gg 1$, 
so $G \left( \lambda \right)$ is always positive for $\lambda \ge 0$.}
\end{figure}

%By introducing the damping term as~\cite{JiHQV,PhysRevA.84.043607} did (from scalar BEC~\cite{PhysRevA.67.033610}), when there is no quadratic Zeeman term, 
By integrating out the $x$ and $y$ directions, the GP equation for a quasi-1D 
spin-$S$ BEC can be written as (see for a detailed derivation Appendix~\ref{quasi_1d_llg_deriv})
\begin{widetext} 
\begin{multline}
\left( i - \Gamma \right) \hbar \frac{\partial \left\{ \Psi \left( z, t \right) \zeta_{\alpha} \left( t \right) \right\}}{\partial t} = %&& \; 
\left\{
- \frac{\hbar^2}{2 m} \frac{\partial^2}{\partial z^2} + V \left( z \right) 
+ \frac{c_0}{2 \pi l_{\perp}^2} n \left( z, t \right) 
\right\} \Psi \left( z, t \right) \zeta_{\alpha} \left( t \right) 
\\
%&& 
+ \hbar 
\left\lbrack
- \boldsymbol{b} 
+ S 
\left\{ 
\boldsymbol{M} \left( t \right) 
- 3 M_z \left( t \right) \boldsymbol{e}_z 
\right\} 
P_{dd} \left( z, t \right) %\qq what is P_{dd} \qq or should it be V_{dd} \qq \Rightarrow It is defined right below: See Eq.(13).
\right\rbrack 
\cdot \left\{ \sum_{\beta = -S}^{S} \left( \boldsymbol{\hat{f}} \right)_{\alpha, \beta} \Psi \left( z, t \right) \zeta_{\beta} \left( t \right) \right\}
\\
%&& 
+ \sum_{k = 1}^{S} \frac{c_{2k}}{2 \pi l_{\perp}^2} n \left( z, t \right) \sum_{\nu_1, \nu_2, \cdots, \nu_k = x, y, z} 
S M_{\nu_1, \nu_2, \cdots, \nu_k} \left( t \right) 
\left\{
\sum_{\beta = -S}^{S} \left( \hat{f}_{\nu_1} \hat{f}_{\nu_2} \cdots \hat{f}_{\nu_k} \right)_{\alpha, \beta} 
\Psi \left( z, t \right) \zeta_{\beta} \left( t \right) 
\right\}, 
\\
\label{quasi_1d_gp_diss_maintext}
\end{multline}
%\end{widetext}
where we defined the two functions 
\begin{align}
M_{\nu_1, \nu_2, \cdots, \nu_k} \left( t \right) \coloneqq \frac{1}{S} \sum_{\alpha, \beta = -S}^{S} \zeta_{\alpha}^{\dagger} \left( t \right) \left( \hat{f}_{\nu_1} \hat{f}_{\nu_2} \cdots \hat{f}_{\nu_k} \right)_{\alpha, \beta} \zeta_{\beta} \left( t \right) , %\nonumber\\
\label{multiple_Mk}
\end{align} 
\begin{align}
P_{dd} \left( z, t \right) &\coloneqq 
\frac{c_{dd}}{2 \hbar  l_{\perp}^3} 
\int_{- \infty}^{\infty} d z' \; 
n \left( z', t \right) 
\left\{ 
G \left( \frac{\left\vert z - z' \right\vert}{l_{\perp}} \right)
 %\right.\nonumber\\& & \left.
- \frac{4}{3} \delta \left( \frac{z - z'}{l_{\perp}} \right) 
\right\}, 
%\nonumber\\
\label{Q_def}
\end{align}
\end{widetext}
with the axial density $n \left( z, t \right) \coloneqq 
\int d^2 \rho \; 
\left\vert \psi \left( \boldsymbol{r}, t \right) \right\vert^2 = \left\vert \Psi \left( z, t \right) \right\vert^2 e^{- 2 \Gamma \omega_{\perp} t / \left( 1 + \Gamma^2 \right)}$,  
where $\int d^2 \rho \coloneqq \int_{- \infty}^{\infty} d x \int_{- \infty}^{\infty} d y$.
Finally, the function $G$ appearing in $P_{dd}$ is defined as 
%Here, 
\begin{eqnarray}
G \left( \lambda \right) \coloneqq \sqrt{\frac{\pi}{2}} \left( \lambda^2 + 1 \right) e^{\lambda^2 / 2} \textrm{Erfc} \left( \frac{\lambda}{\sqrt{2}} \right) - \lambda .
\label{F_defin}
\end{eqnarray}
We plot $G \left( \lambda \right)$ as a function of $\lambda$ in Fig.~\ref{G-plot}. 
Eq.\eqref{quasi_1d_gp_diss_maintext} represents our starting point for
analyzing the dynamics of magnetization.  We will now proceed to show how it
leads to the LLG equation and the Stoner-Wolfarth model.

%From Eq.~\eqref{quasi_1d_wavefunc}, $\int d \boldsymbol{\rho} \; \left\vert \psi \left( \boldsymbol{r}, t \right) \right\vert^2 = \left\vert \Psi \left( z, t \right) \right\vert^2 e^{- 2 \Gamma \omega_{\perp} t / \left( 1 + \Gamma^2 \right)}$. Hence, if we renormalize $\psi \left( \boldsymbol{r}, t \right)$ every time so that $\int d \boldsymbol{r} \; \left\vert \psi \left( \boldsymbol{r}, t \right) \right\vert^2 \coloneqq N$ becomes constant in time, $\left\vert \Psi \left( z, t \right) \right\vert^2 e^{- 2 \Gamma \omega_{\perp} t / \left( 1 + \Gamma^2 \right)}$ should be constant in time, which we will call $n \left( z \right)$. Hence, $P \left( z, t \right)$ becomes constant in time (which we will denote as $P \left( z \right)$) when we renormalize $\psi \left( \boldsymbol{r}, t \right)$ every time.

\section{
\label{Lagrangian_discussion}
Effective Langrangian description} 
%for magnetization dynamics} %constant axial density}
%To study dynamics when there is dipole-dipole interaction, w
To provide a concise phase space picture of the condensate magnetization dynamics,   
we discuss in this section a collective coordinate Lagrangian appropriate to our system.
%calculate effective Lagrangian of our system. To do so, we have to
%calculate dipole-dipole interaction term $V_d$.

Let $\boldsymbol{M} \left( t \right) \coloneqq \boldsymbol{d} \left( z, t \right) / d \left( z, t \right)$ where the magnetization 
$\boldsymbol{d} \left( z, t \right)$ is defined in Eq.~\eqref{d_def}. Explicitly, the local  magnetization direction reads 
$\boldsymbol{M} \left( t \right) = \left( \sin \theta \left( t \right) \cos \phi \left( t \right), \sin \theta \left( t \right) \sin \phi \left( t \right), \cos \theta \left( t \right) \right)$.
Then, from Eqs.~\eqref{F_expect_val} and~\eqref{psi_sq_explicit}, 
$\boldsymbol{F} \left( \boldsymbol{r}, t \right) = S\boldsymbol{M} \left( t \right) \left\vert \psi \left( \boldsymbol{r}, t \right) \right\vert^2$ 
and one obtains (see for a detailed derivation Appendix~\ref{quasi_1d_llg_deriv})
\begin{eqnarray}
\frac{\partial \boldsymbol{M}}{\partial t} = && \; 
\boldsymbol{M} \times 
\left\{ 
\boldsymbol{b} + S 
\Lambda'_{dd} \left( t \right) 
M_{z} \boldsymbol{e}_{z} 
\right\} 
- \Gamma \boldsymbol{M} \times 
\frac{\partial \boldsymbol{M}}{\partial t}, \quad\quad\;
\label{SW_LLG_st1}
\end{eqnarray}
where the renormalized interaction function $\Lambda'_{dd} \left( t \right)$ reads 
\begin{eqnarray}
\Lambda'_{dd} \left( t \right) = && \; \frac{3}{N \left( t \right)} 
\int_{- \infty}^{\infty} d z \; 
n \left( z, t \right) P_{dd} \left( z, t \right),
\label{lambdap_def}
\end{eqnarray} 
and $N \left( t \right) \coloneqq 
%\int d \boldsymbol{r} \; 
\int d^3 r \; 
\left\vert \psi \left( \boldsymbol{r}, t \right) \right\vert^2 = 
\int_{- \infty}^{\infty} d z \; 
n \left( z, t \right)$.
%cf.~the derivation contained in Appendix~\ref{quasi_1d_llg_deriv}.
From Eqs.~\eqref{Veff_Vd_connec}, \eqref{V_eff_real_space}, and~\eqref{Q_def}, $\Lambda'_{dd} \left( t \right)$ is connected to the dipole-dipole interaction contribution $V_{dd} \left( t \right)$ by 
\begin{eqnarray}
V_{dd} \left( t \right) = && \; 
\frac{3}{2} \hbar S^2 
\left\{ 
\sin^2 \theta \left( t \right) - \frac{2}{3} 
\right\} 
\int_{- \infty}^{\infty} d z \; 
n \left( z, t \right) P_{dd} \left( z, t \right) 
\nonumber\\
= && \; \frac{\hbar}{2} S^2 N \left( t \right) \Lambda'_{dd} \left( t \right) 
\left\{ 
\frac{1}{3} - \cos^2 \theta \left( t \right) 
\right\} . 
\label{dipole_interaction_explicit_gen}
\end{eqnarray}
We note that in order to obtain the effective quasi-1D dipolar interaction \eqref{dipole_interaction_explicit_gen}, 
we did not use, in distinction to Ref.~\cite{Giovanazzi2004}, 
any simplifying approximation. A detailed derivation is provided in 
Appendix~\ref{G_deriv}.

Eq.\eqref{SW_LLG_st1} is the 
%Landau-Lifshitz-Gilbert 
LLG equation with the external magnetic field in $z$-direction modified by the magnetization in $z$-direction due to the dipole-dipole interaction.  The corresponding term in units of magnetic field,  $\hbar S \Lambda'_{dd} \left( t \right) M_{z} \boldsymbol{e}_{z} / \left( g_{F} \mu_{B} \right)$, can be seen as an additional magnetic field that is itself proportional to the magnetization in $z$-direction, and which leads to an additional nonlinearity in the LLG equation.

From Eqs.~\eqref{Q_def} and~\eqref{lambdap_def}, to get how $\Lambda'_{dd} \left( t \right)$ depends on time $t$, 
one has to calculate the double integral 
\begin{align}
\int \!\! d z \!\! \int \!\! d z' %&& \; 
n \left( z, t \right) n \left( z', t \right) 
\left\{ 
G \left( \frac{\left\vert z - z' \right\vert}{l_{\perp}} \right) \!
- \frac{4}{3} \delta \left( \frac{z - z'}{l_{\perp}} \right) 
\right\} . 
%\nonumber\\
\label{double_integ_n_n}
\end{align}
To achieve a simple physical picture, 
we assume that $n \left( z, t \right)$ does not depend on time $t$ within the time range we are interested in. 
Then we may write $\Lambda'_{dd} \left( t \right) = \Lambda'_{dd}$. 
The lifetime of a typical dipolar %spinor 
BEC with large atomic magnetic dipole moments 
such as ${}^{164} \textrm{Dy}$ \cite{Lev}, ${}^{162} \textrm{Dy}$ and ${}^{160} \textrm{Dy}$ 
\cite{LevNJP}, or ${}^{166} \textrm{Er}$ \cite{Er_quasi_1D_exp}
is of the order of seconds.
%, and the number of condensed atoms in the central core region is almost constant within 10 ms in the spinor gas experiment of Ref.~\cite{Er_quasi_1D_exp}. 
Since taking into account the time dependence of $n \left( z, t \right)$ generally requires a numerical solution of Eq.~\eqref{quasi_1d_gp_diss_maintext}, we here consider the case where $n \left( z, t \right)$ is constant in time $t$ as in~\cite{PhysRevA.84.043607}, to predominantly extract the effect of magnetic dipole-dipole interaction {\it per se}.

We also neglect the possible effect of magnetostriction.  The latter effect, amounting to  
a distortion of the aspect ratio of the condensate in a harmonic trap as a function of the angle of the external magnetic field with the symmetry axis of the trap, was measured in a condensate of Chromium atoms \cite{stuhler_magnetostriction_2007} (with a magnetic moment of $6\,\mu_\text{B}$). The magnetostriction effect in that experiment was of the order of 10\%. For alkali atoms with spin-1 the effect should be a factor $6^2$  
smaller.  In addition, theoretical analyses in the Thomas-Fermi limit show that magnetostriction in harmonic traps becomes particularly small for very small or very large asymmetries of the trap \cite{giovanazzi_ballistic_2003,sapina_ground-state_2010}.   

More specifically, Ref.~\cite{PhysRevLett.89.130401} has shown that magnetostriction is due to the force induced by the dipole-dipole mean-field potential $\Phi_{dd} \left( \boldsymbol{r}, t \right)$. 
In Appendix~\ref{magnetostriction-appendix}, we apply the approach of \cite{PhysRevLett.89.130401} to a dipolar spinor BEC. 
From Eqs.~\eqref{lambdap_def},~\eqref{dipole_interaction_explicit_gen},~\eqref{V_d_real_eff}, and~\eqref{Phi-dd-def}, 
$\Lambda'_{dd} \left( t \right)$ contains $\Phi_{dd} \left( z, t \right)$ [the quasi-1D form of $\Phi_{dd} \left( \boldsymbol{r}, t \right)$ defined in Eq.~\eqref{Phi-dd-def}] 
by 
\begin{multline}
S^2 \left\{ 1 - 3 M_z^2 \left( t \right) \right\} N \left( t \right) \hbar \Lambda'_{dd} \left( t \right) 
\\ = 
%&& \; 
3 
\int_{- \infty}^{\infty} d z \; 
n \left( z, t \right) 
\Phi_{dd} \left( z, t \right) .
%\nonumber\\
\label{Phi-dd-Lambda-dd-connec}
\end{multline}
Hence, our LLG-type equation in Eq.~\eqref{SW_LLG_st1} 
effectively contains the dipole-dipole mean-field potential which causes magnetostriction 
and the form of Eq.~\eqref{SW_LLG_st1} itself will not be changed whether the effect of magnetostriction is large or not. Only the value of $\Lambda'_{dd} \left( t \right)$ will be changed because  magnetostriction changes  the integration domain. 
Furthermore, we show %\qq in addition
in Appendix~\ref{magnetostriction-appendix} that 
%Newly-Added-by-Shinn
for our quasi-1D system, 
the effect of magnetostriction is smaller in a box trap than in harmonic trap.
In fact, for the box trap, this effect can be neglected if 
%Newly-Added-by-Shinn
$L_z / l_{\perp}$ is sufficiently large. 
Thus, we may neglect the effect of magnetostriction under suitable limits for both box and harmonic traps.

%Hence, $K$ in Stoner-Wohlfarth model is {
%$N \hbar \Lambda_{d} \left( L_z / l_{\perp} \right) / 2$, }
%which is positive since $L_z \gg l_{\perp}$ in 
%{quasi-1D } space in $z$ axis.

To get a simple physical idea of the dynamical behavior of our system, let us, for now, assume that there is no damping, $\Gamma = 0$. When the external magnetic field is chosen to lie in the $x-z$ plane, $\boldsymbol{B}=\left( B_x, 0, B_z \right)$, Eq.~\eqref{SW_LLG_st1} becomes 
\begin{eqnarray}
\frac{d \theta}{d t} = && \; b_x \sin \phi,
\nonumber\\
\frac{d \phi}{d t} = && \; b_x \cot \theta \cos \phi - b_z - S \Lambda'_{dd} \cos \theta.
\label{th_ph_simple_eq}
\end{eqnarray} 
where we already defined the Larmor frequency vector $\boldsymbol{b} = g_F \mu_B \boldsymbol{B} / \hbar$ below Eq.~\eqref{GP_SW_q}. 

By using the Lagrangian formalism introduced in~\cite{doi:10.1119/1.4709188}, the Lagrangian $L$ of this system then fulfills 
\begin{eqnarray}
\frac L\hbar = && \; \dot{\phi} \cos \theta + b_x \sin \theta \cos \phi + b_z \cos \theta 
+ \frac{S}{4} \Lambda'_{dd} \cos \left( 2 \theta \right) , 
\nonumber\\
\label{Lagrangian_simple}
\end{eqnarray}
where $\dot{\phi} = d \phi / d t$.
The equations of motion are 
%From Eq.~\eqref{Lagrangian_simple}, one can get 
\begin{eqnarray}
\frac1\hbar \frac{\partial L}{\partial \theta} = && \; - \dot{\phi} \sin \theta + b_x \cos \theta \cos \phi - b_z \sin \theta 
- \frac{S}{2} \Lambda'_{dd} \sin \left( 2 \theta \right),
\nonumber\\
\frac{\partial L}{\partial \dot{\theta}} = && \; 0,
\quad 
\frac1\hbar\frac{\partial L}{\partial \phi} = - b_x \sin \theta \sin \phi, 
\quad
\frac1\hbar\frac{\partial L}{\partial \dot{\phi}} = \cos \theta.
\quad\quad
\end{eqnarray}
One easily verifies that Eq.~\eqref{Lagrangian_simple} is indeed the Lagrangian which gives Eqs.~\eqref{th_ph_simple_eq}. Let $p_{\xi}$ be the conjugate momentum of the coordinate $\xi$. Since $p_{\theta} = 0$ and $p_{\phi} = \hbar \cos \theta$ ($\hbar$ times the $z$ component of 
$\boldsymbol M$), the Hamiltonian $H$ is given by 
%\begin{multline}
\begin{align}
H = - b_x \sqrt{\hbar^2 - p_{\phi}^2} \cos \phi - b_z p_{\phi} 
+ \frac{\hbar^2 - 2 p_{\phi}^2}{4 \hbar} S \Lambda'_{dd} . 
%\\
\label{Ham_const_den_1D}
%\end{multline}
\end{align}
Note that the energy 
$\tilde{E} \coloneqq H - \hbar S \Lambda'_{dd} / 4$ 
is conserved. Hence, if we put $p_{\phi} = \left( p_{\phi} \right)_{\rm in}$ and $\phi = \pi / 2$ at some time $t = t_0$, 
$\tilde{E} = - b_z \left( p_{\phi} \right)_{\rm in} - S \Lambda'_{dd} \left( p_{\phi} \right)_{\rm in}^2 / 2\hbar$. 
We can then express $\phi$ as a function of $p_{\phi}$ as
\begin{eqnarray}
\cos \phi = && \; - \frac{\tilde{E} + b_z p_{\phi} + \frac{1}{2 \hbar} S \Lambda'_{dd} p_{\phi}^2}{b_x \sqrt{\hbar^2 - p_{\phi}^2}} 
\nonumber\\
= && \; \left\{ \left( p_{\phi} \right)_{\rm in} - p_{\phi} \right\} \frac{b_z + S \Lambda'_{dd} \frac{\left( p_{\phi} \right)_{\rm in} + p_{\phi}}{2\hbar }}{b_x \sqrt{\hbar^2- p_{\phi}^2}}. \quad\quad
\label{phase_diag_const_exp}
\end{eqnarray}
The canonical momentum $p_{\phi}$ remains the initial $\left( p_{\phi} \right)_{\rm in}$ when $b_x = 0$, implying that $\theta$ does not change when $b_x = 0$, consistent with Eqs.~\eqref{th_ph_simple_eq}. 
If $\left\vert b_x \right\vert$ is larger than $\left\vert b_z \pm S \Lambda'_{dd} \right\vert$, we can have $p_{\phi} \neq \left( p_{\phi} \right)_{\rm in}$ with $\left\vert \cos \phi \right\vert \le 1$, which allows for the switching process of the magnetization.  
Below a threshold value of $|b_x|$ that depends on $b_z$ and $S \Lambda'_{dd}$, $ p_{\phi}$ has to remain constant for Eq.~\eqref{phase_diag_const_exp} to be satisfied, which corresponds to simple magnetization precession about the $z$ axis.  
%Also, to make $p_{\phi} \neq \left( p_{\phi} \right)_{\rm in}$, $\left\vert b_x \right\vert$ should be larger than $\left\vert b_z \pm S \Lambda_{dd} \left( L_z / l_{\perp} \right) \right\vert$, so that $\left\vert \cos \phi \right\vert \le 1$. Else $p_{\phi}$ would remain constant, which corresponds to simple magnetization precession about the $z$ axis. 

\begin{figure} [t]
\centering
\subfigure{%\label{phi_p_phase}
\includegraphics[width=0.4\textwidth]{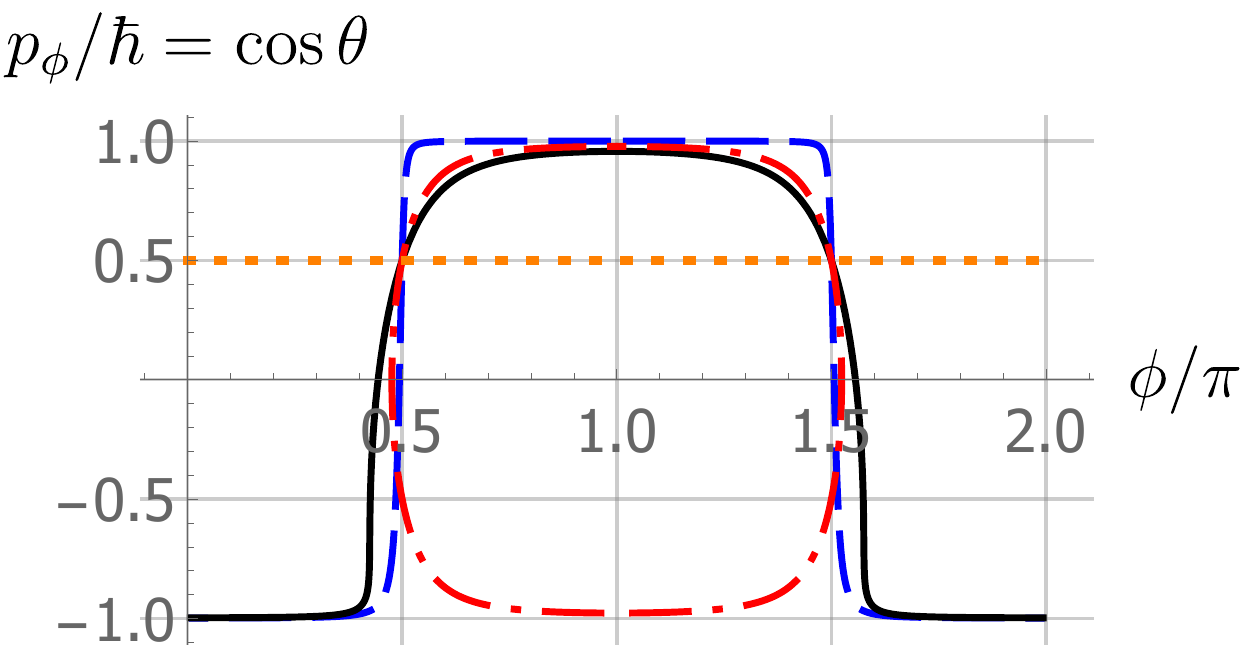}}
{$\left( p_{\phi} \right)_{\rm in} = \hbar/2$
and $\phi_{\rm in} = \pi / 2$ }
\subfigure{%\label{phi_p_phase_neg}
\includegraphics[width=0.4\textwidth]{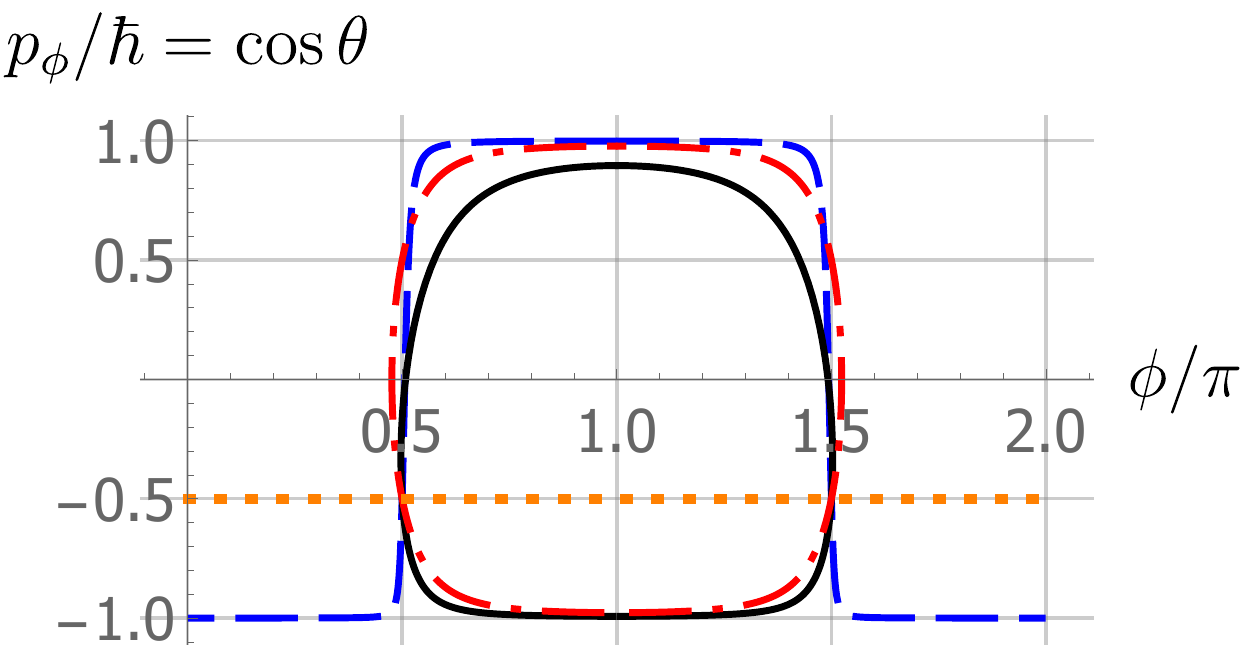}}
{$\left( p_{\phi} \right)_{\rm in} = -\hbar/2$and $\phi_{\rm in} = \pi / 2$}
\caption{\label{phi_p_phase} $p_{\phi}/\hbar$ vs $\phi / \pi$ 
when $\Gamma = 0$ (no dissipation), with initial values $\left( p_{\phi} \right)_{\rm in}$ 
and $\phi_{\rm in}$ (initial value of $\phi$) 
as shown. 
(1) Dashed blue: $b_z / b_x = 0.2$ and 
$\left\vert b_z \right\vert \gg S \Lambda'_{dd} $. 
(2) Black line: $b_z / b_x = 0.2$ and 
$S \Lambda'_{dd} / b_x = 0.6$. 
(3) Dash-Dotted red: $S \Lambda'_{dd} / b_x = 0.6$ and $\left\vert b_z \right\vert \ll S \Lambda'_{dd}$. 
(4) Dotted orange horizontal line: $b_x = 0$.}
\end{figure}

When $p_{\phi}$ is a function of time, there are two important cases:
\begin{eqnarray}
\textrm{(a) } & & \left\vert b_z \right\vert \gg S \Lambda'_{dd}: \;
\cos \phi = \frac{b_z}{b_x} \frac{\left( p_{\phi} \right)_{\rm in} - p_{\phi}}{\sqrt{\hbar^2 - p_{\phi}^2}}, 
\nonumber\\
\textrm{(b) } & &  \left\vert b_z \right\vert \ll S \Lambda'_{dd}: \; 
\cos \phi = \frac{S \Lambda'_{dd}}{2 b_x} \frac{\left( p_{\phi} \right)_{\rm in}^2 - p_{\phi}^2}{\hbar \sqrt{\hbar^2 - p_{\phi}^2}}. 
%\nonumber\\
\label{phi_p_eqs}
\end{eqnarray} 
We plot the corresponding phase diagrams ($\theta$ vs $\phi$) in Fig.~\ref{phi_p_phase}.
% and~\ref{phi_p_phase_neg}.

Let $\left( p_{\phi} \right)_{\rm in} = \hbar\cos \theta_{\rm in}$, $b_x = b \sin \theta_0$, and $b_z = b \cos \theta_0$. When case (a) holds   
$\left\vert b_z \right\vert \gg S \Lambda'_{dd}$, 
one concludes that $\cos \theta_0  \cos \theta + \sin \theta_0 \sin \theta \cos \phi = \cos \theta_0 \cos \theta_{\rm in}$, which is constant. 
%where $\left( p_{\phi} \right)_{\rm in} = \cos \theta_{\rm in}$. 
Since $\boldsymbol{d} \cdot \boldsymbol{b} = d b \left( \cos \theta_0 \cos \theta + \sin \theta_0 \sin \theta \cos \phi \right)$,  
%where $b_x = b \sin \theta_0$ and $b_z = b \cos \theta_0$, 
in case (a) the magnetization $\boldsymbol{d}$ precesses 
around the external magnetic field $\boldsymbol{B}$,  
%if 
%{$\left\vert b_z \right\vert \gg \Lambda_{d} \left( L_z / l_{\perp} \right)$, }
as expected.
When (b) holds, SW switching can occur, to the description of which we proceed in the following.

\section{\label{connec_to_SW}Connection to Stoner-Wohlfarth Model}
The phenomenological SW model can be directly read off from the equations in the preceding section.  
From Eq.~\eqref{Ham_const_den_1D}, $\tilde{H} \coloneqq H + \hbar S \Lambda'_{dd} / 4$ 
is given by 
\begin{eqnarray}
\frac{\tilde{H}}\hbar = - b_x \sin \theta \cos \phi - b_z \cos \theta + \frac{S \Lambda'_{dd}}{2} \sin^2 \theta. \nonumber\\
\label{Ham_cons_den_1D_red}
\end{eqnarray}
Let $\left( b_{\nu} \right)_{\rm cr}$ be the value of $b_{\nu}$ at the stability limit where $\partial \tilde{H} / \partial \theta = 0$ and $\partial^2 \tilde{H} / \partial \theta^2 = 0$. Then one obtains
the critical magnetic fields  
\begin{eqnarray}
&& \left( b_x \right)_{\rm cr} \cos \phi = S \Lambda'_{dd} \sin^3 \theta , 
\quad 
\left( b_z \right)_{\rm cr} = - S \Lambda'_{dd} \cos^3 \theta .
\nonumber\\
\end{eqnarray}
which satisfy the equation
\begin{eqnarray} 
\left\{ \left( b_x \right)_{\rm cr} \cos \phi \right\}^{2/3} + \left( b_z \right)_{\rm cr}^{2/3} = \left\{ S \Lambda'_{dd} \right\}^{2/3} . 
\label{gen-switching-curve}
\end{eqnarray}
We coin the curve in the $(b_x,b_z)$-plane described by Eq.~\eqref{gen-switching-curve} the switching curve,  in accordance with the terminology established in~\cite{hubert1998magnetic}. 
Because $\phi$ changes in time [see Eqs.~\eqref{th_ph_simple_eq} and Fig.~\ref{phi_p_phase}], the switching curve depends in general on the timing of the applied external magnetic fields.
We note that, for $\phi = 0$, Eqs.~\eqref{Ham_cons_den_1D_red} and~\eqref{gen-switching-curve} are identical to the SW energy functional 
\begin{equation}\label{SW_H} 
\frac{H_{\rm SW}}\hbar = - b_x \sin \theta -b_z \cos \theta + K \sin^2 \theta
\end{equation} 
and the SW astroid~\cite{hubert1998magnetic}, respectively,  
%(with suitable scaling)  
if we identify 
$K = S \Lambda'_{dd} / 2$.
%According to~\cite{pitaevskii1959phenomenological,Zaremba1999,PhysRevA.67.033610}, one %may employ the phenomenological damping term in the Gross-Pitaevski\v{\i} (GP) equation, %and its validity for scalar BEC has been shown by numerical simulations in~\cite{PhysRevA.%57.4057}. 

The LLG equation in 
Eq.~\eqref{SW_LLG_st1} has stationary solutions %when
with $\boldsymbol{M}$ 
parallel to the effective magnetic field 
$\hbar \left\{ \boldsymbol{b} + S \Lambda'_{dd} \left( t \right) M_{z} \boldsymbol{e}_{z} \right\} / \left( g_{F} \mu_{B} \right)$.  
Since we set $\boldsymbol{b}$ to lie in the $xz$ plane, 
%lies in $xz$ plane (Note that we set ). Hence, 
$\phi$ will go to zero %as time goes,
for sufficiently large times.  Thus Eq.~\eqref{Ham_cons_den_1D_red} %corresponds
leads 
to the SW
%Stoner-Wohlfarth 
model \eqref{SW_H} %if we include a
  due to the damping term in \eqref{SW_LLG_st1} if $\Gamma > 0$.
% and renormalize wavefunction. 
%Note that, if $n \left( z \right)$ is not constant, then we only need to change $\Lambda \left( L_z / l_{\perp} \right)$ to $\Lambda'$ where 
In ~Appendix \ref{Gamma_tensor},  we demonstrate 
that a more general tensorial damping coefficient $\Gamma$  
introduces additional terms on the right-hand side of the LLG equation 
\eqref{SW_LLG_st1}, which involve {\it time derivatives}.
While these will thus not affect the SW
%Stoner-Wohlfarth 
phenomenology, 
which results from the steady states as function of the applied magnetic fields, 
%,   that we discuss in the following, 
and which is thus governed by the vanishing (in the stationary limit)  of the 
first term on the right-hand side of the LLG equation,
they affect the detailed relaxation dynamics of the magnetization
and its time scales. These deviations can hence can be used to probe deviations from assuming a single scalar $\Gamma$. 

%By integrating out $z$ axis and renormalizing wavefunction, Eq.~\eqref{SW_LLG_st1} can be written as 
%\begin{eqnarray}
%\frac{\partial \boldsymbol{M}}{\partial t} = && \; 
%\boldsymbol{M} \times 
%\left( 
%\boldsymbol{b} + S \Lambda' M_{z} \boldsymbol{e}_{z} 
%\right) 
%- \Gamma \boldsymbol{M} \times 
%\frac{\partial \boldsymbol{M}}{\partial t}. 
%\label{SW_LLG_st2}
%\end{eqnarray}

%For ferromagnetic metals, \cite{PhysRevB.74.144405} derived the phenomenological damping %term in 
%LLG 
%Landau-Lifshitz-Gilbert 
%equation (which has similar form as Eq.~\eqref{SW_LLG_st1}) by quantum kinetic equation. %Hence, in principle one may derive $\Gamma$. However, it is beyond our subject and we will %not do that in this paper.

  Before we move on to the next section, %\qq to
  we %\qq
  show the characteristic behavior of 
$\Lambda'_{dd}$ defined in Eq.~\eqref{lambdap_def}, 
% \qq we calculate it by assuming
for a box-trap scenario defined by 
$n \left( z, t \right) = N / \left( 2 L_z \right)$ for $- L_z \le z \le L_z$ and $n \left( z, t \right) = 0$ otherwise ($N$ is number of particles). 

We stress that due to the finite size of the trap along the ``long" $z$ direction, 
in variance with the  Hohenberg-Mermin-Wagner theorem holding for infinitely 
extended systems in the thermodynamic limit,  a quasi-1D BEC can exist also at finite temperatures~\cite{PhysRevLett.89.280402}. This remains true up to a ratio of its proper 
length to the de-Broglie wavelength \cite{JLTP}, beyond which %\qq length
strong 
phase fluctuations set in \cite{Dettmer}. 
In fact, these strongly elongated quasi-1D BECs at finite temperature have been first realized already long ago, cf., e.g.~\cite{PhysRevLett.87.130402}.  

%as is implied in Eq.(13) of~ \cite{PhysRevLett.89.280402}. 
%Thus, considering box trap scenario can be justified and it will illustrate characteristics of $%\Lambda'_{dd}$. 
For the box trap, $\Lambda'_{dd} = \Lambda_{dd} \left( L_z / l_{\perp} \right)$ where 
\begin{equation}
\Lambda_{dd} \left( \lambda \right) = \frac{3 N c_{dd}}{2 \hbar l_{\perp}^3} \frac{1}{\lambda}  
\left\{ 
\int_{0}^{2 \lambda} d v \; 
\left( 1 - \frac{v}{2 \lambda} \right) G \left( v \right) - \frac{2}{3} 
\right\} . 
\label{dipole_interaction_explicit_Lambda}
\end{equation} 
From Eq.~\eqref{F_defin},  
$G \left( v \right) \simeq 2 / v^3 + \ord\left( v^{-5} \right)$ for 
%$v \rightarrow \infty$, so that 
$v \gg 1$, so that 
\begin{equation} 
\Lambda_{dd} \left( \lambda \right)  \simeq \frac{N c_{dd}}{2 \hbar l_{\perp}^3} \frac{1}{\lambda}  
\qquad \mbox{for $\lambda = \frac{L_z}{l_\perp} \gg 1$}.
\end{equation}
%$\bar{\Lambda}_{dd} \left( \lambda \right) \coloneqq 2 \hbar l_{\perp}^3 
%$\Lambda_{dd} \left( \lambda \right) 
%/ \left( 3 N c_{dd} \right)
%\asymp 
%\propto 1/\lambda $ for $\lambda \rightarrow \infty$. 
Hence $\Lambda_{dd} \left( \lambda \right)$ is a slowly decreasing function of the cigar's aspect ratio $\lambda$ (keeping everything else fixed). We will see below that for the parameters of experiments such as  \cite{Er_quasi_1D_exp}, the effective magnetic field due to dipolar interactions %in the gas easily 
greatly exceeds the externally applied magnetic fields 
(in the range relevant for SW switching to be observed) \footnote{This fact is equivalent to the critical dimensionless magnetization in Eq.~\eqref{def_M_c} being always much less than unity.}.

\section{\label{numerical_res}
Analytical Results for Axially Directed External Magnetic Field}
Without dissipation, when $b_x = 0$, $p_{\phi} = \hbar\cos \theta = \hbar M_z$ is rendered constant; see Eq.~\eqref{th_ph_simple_eq}. However, in the presence of dissipation, $M_z$ changes in time even if $b_x = 0$. By employing this change, we propose an experimental method to measure $\Gamma$. 

For simplicity, we will assume that 
the number density is constant in time (also see section ~\ref{Lagrangian_discussion}) and 
the external magnetic field %\qq is pointing
points along the $z$ direction, 
$\boldsymbol{B} = B_z \boldsymbol{e}_z$.  
Let a critical (see for a detailed discussion below) value of the magnetization be 
\begin{equation}
\left( M_z \right)_{\rm cr} \coloneqq - \frac{b_z}{S \Lambda'_{dd} }. 
\label{def_M_c}
\end{equation}  
Then Eq.~\eqref{SW_LLG_st1} can be written as 
\begin{eqnarray}
\frac{\partial \boldsymbol{M}}{\partial t} = && \; 
S \Lambda'_{dd} \boldsymbol{M} \times \boldsymbol{e}_{z} 
\left\{ 
M_{z} - \left( M_z \right)_{\rm cr}
\right\} 
- \Gamma \boldsymbol{M} \times 
\frac{\partial \boldsymbol{M}}{\partial t}
\nonumber\\
= && \; 
\boldsymbol{M} \times \boldsymbol{e}_{z} 
\left( 
b_z + S \Lambda'_{dd} M_{z} 
\right) 
- \Gamma \boldsymbol{M} \times 
\frac{\partial \boldsymbol{M}}{\partial t}. \quad\quad
\label{SW_LLG_st3}
\end{eqnarray}
Since $\boldsymbol{M} \cdot \frac{\partial \boldsymbol{M}}{\partial t} = 0$, by taking the cross product with $\boldsymbol{M}$ on both sides of Eq.~\eqref{SW_LLG_st1}, one can derive an expression for $\boldsymbol{M} \times \frac{\partial \boldsymbol{M}}{\partial t}$:  
\begin{eqnarray}
\frac{\partial M_z}{\partial t} = && \; 
- \frac{\Gamma S \Lambda'_{dd}}{1 + \Gamma^2}  
\left\{ 
M_{z} - \left( M_z \right)_{\rm cr}
\right\} 
\left( M_{z}^2 - 1 \right)
\nonumber\\
= && \; 
- \frac{\Gamma}{1 + \Gamma^2}  
\left( 
b_z + S \Lambda'_{dd} M_{z} 
\right) 
\left( M_{z}^2 - 1 \right). 
\label{SW_LLG_Mz_eq}
\end{eqnarray}
Since $\boldsymbol{M}$ is the scaled magnetization, $\left\vert \boldsymbol{M} \right\vert = 1$ with a condensate.
%in mean field. % as long as the number of condensed atoms is not zero. 
Hence, $-1 \le M_z \le 1$. Also, according to the discussion below Eq.~\eqref{Ham_cons_den_1D_red}, the generally positive SW coefficient 
(with units of frequency) $K$ is $S \Lambda'_{dd} / 2$.
% and $K$ is generally positive in SW model. 

From Eq.~\eqref{SW_LLG_Mz_eq}, 
for time-independent $\Lambda'_{dd}$, 
one concludes that there are three time-independent solutions,  
%(reducing to two solutions if $\Lambda'_{dd}$ depends on time $t$ \qq meaning what\qq): 
% $\Rightarrow$ If $\Lambda'_{dd}$ depends on time $t$, $\left( M_z \right)_{\rm cr}$ also depends on $t$. Then $M_z = \left( M_z \right)_{\rm cr}$ is NOT time-independent solution.
$M_z = \left( M_z \right)_{\rm cr}$ and $M_z = \pm 1$. 
For a box-trapped BEC and constant number density, 
$\Lambda'_{dd} = \Lambda_{dd}$ which is always positive in the quasi-1D limit 
(cf.~Eq.~\eqref{dipole_interaction_explicit_Lambda} and the discussion following it). 
For some arbitrary physical quasi-1D trap potential, in which the number density is not constant in space, from Eqs.~\eqref{Q_def},~\eqref{lambdap_def}, and Fig.~\ref{G-plot}, one can infer that $\Lambda'_{dd} > 0$, due to the fact that the quasi-1D number density $n \left( z, t \right) > 0$, $n \left( z, t \right)$ has its maximum value near $z = 0$ for a symmetric trap centered there,  and then $G \left( \lambda \right)$ also has its maximum value near $\lambda = 0$.
%(see Fig.~\ref{lambda-plot})  
 %We thus consider $\Lambda'_{dd} > 0$. 
Then, if $\left\vert \left( M_z \right)_{\rm cr} \right\vert < 1$, 
%If $\Gamma S \Lambda'_{d} > 0 $ 
%(generally true within the parameter regime of interest).
%Finally, 
$M_z = \left( M_z \right)_{\rm cr}$ is an unstable solution 
and $M_z = \pm 1$ are stable solutions. 
When $\left\vert \left( M_z \right)_{\rm cr} \right\vert < 1$ and 
$-1 < M_z < \left( M_z \right)_{\rm cr}$, $M_z$ goes to $-1$. 
Likewise, $M_z$ goes to 1 when $\left( M_z \right)_{\rm cr} < M_z < 1$. 
This bifurcation does not occur if $\left\vert \left( M_z \right)_{\rm cr} \right\vert> 1$. 
For simplicity, we assume that $\left\vert \left( M_z \right)_{\rm cr} \right\vert < 1$.
This is the more interesting case due to the possibility of a bifurcation of stable solutions 
leading to SW switching.

Let $\left( M_z \right)_{\rm in}$ be the value of $M_z$ at $t = 0$. 
%To obtain the analytical solution of Eq.~\eqref{SW_LLG_Mz_eq}, we assume that the number density in $z$ axis $n \left( z, t \right)$ is constant in time $t$  so that also $\Lambda'_{dd}$ is constant in time. Typical lifetimes of dipolar spinor BECs are of the order of seconds, and the number of condensed atoms of the central core is almost constant within 10 ms for the experiment conducted by~\cite{Er_quasi_1D_exp}. Hence, if $M_z$ converges to its asympotics within 10 ms for the parameters of the latter experiment, our assumption can be justified.
%, similar to~\cite{PhysRevA.84.043607}, which  assumed a number of condensed atoms %remaining roughly constant even in the presence of dissipation.
The analytic solution of Eq.~\eqref{SW_LLG_Mz_eq} satisfies 
%\begin{widetext}
\begin{eqnarray}
t & = &  \frac{1 + \Gamma^2}{\Gamma S \Lambda'_{dd}} 
\left\lbrack 
\frac{1}{\left\{ \left( M_z \right)_{\rm cr}\right\}^2 - 1} 
\ln \left\{
\frac{\left( M_z \right)_{\rm in} - \left( M_z \right)_{\rm cr}}{M_z - \left( M_z \right)_{\rm cr}}
\right\} 
 \right. \nonumber\\& & \left. 
- \frac{1}{2 \left\{ 1 - \left( M_z \right)_{\rm cr} \right\}} 
\ln \left\{ 
\frac{1 - M_z}{1 - \left( M_z \right)_{\rm in}} 
\right\}  \right. \nonumber\\& & \left. 
+ \frac{1}{2 \left\{ 1 + \left( M_z \right)_{\rm cr} \right\}} 
\ln \left\{ 
\frac{1 + \left( M_z \right)_{\rm in}}{1 + M_z}
\right\} 
\right\rbrack 
\nonumber\\
& = & \frac{1 + \Gamma^2}{\Gamma} 
\left\lbrack 
\frac{S \Lambda'_{dd}}{b_z^2 - \left( S \Lambda'_{dd} \right)^2} 
\ln \left\{
\frac{b_z + S \Lambda'_{dd} \left( M_z \right)_{\rm in}}{b_z + S \Lambda'_{dd} M_z}
\right\} 
 \right. \nonumber\\& & \left. 
- \frac{1}{2 \left( b_z + S \Lambda'_{dd} \right)} 
\ln \left\{ 
\frac{1 - M_z}{1 - \left( M_z \right)_{\rm in}} 
\right\} \right. \nonumber \\& &\left.
-\frac{1}{2 \left( b_z - S \Lambda'_{dd} \right)} 
\ln \left\{ 
\frac{1 + \left( M_z \right)_{\rm in}}{1 + M_z}
\right\} 
\right\rbrack .
\label{Mz_analytic_sol}
\end{eqnarray}
%\end{widetext}
The above equation tells us that, if $\left( M_z \right)_{\rm in} \neq \left( M_z \right)_{\rm cr}$ 
and $\left( M_z \right)_{\rm in} \neq \pm 1$, 
$M_z$ goes to its stable time-independent solution ($\left\vert M_z \right\vert = 1$) at time $t = \infty$. 
Thus, we define a {\it critical switching time} 
$t_{\rm cr}$ to be the time when $\left\vert M_z \right\vert = 0.99$. 
Also, note that the form of LLG equation (Eq.~\eqref{SW_LLG_st3}) does not change whether BEC is confined in a quasi-1D, quasi-2D, or a three-dimensional geometry. 
This is because one can find a connection between 
$\Lambda'_{dd}$ and the effective dipole-dipole-interaction potential %\qq hyphen added
$V_{\rm eff}$,  %like Eqs.~\eqref{Q_def} and~\eqref{lambdap_def}, 
so one can measure $\Gamma$ even if the BEC is effectively confined in a space with dimension higher than one, 
% (``higher'' means ``higher than quasi-1D'') 
using Eq.~\eqref{Mz_analytic_sol}.
% as long as $\Gamma$ do not depend on spin indices.

We point out, in particular, that $t_{\rm cr}$ is inversely proportional to 
$\Lambda'_{dd}$. 
Hence, for a constant density quasi-1D BEC 
confined between $- L_z \le z \le L_z$, 
$\Lambda'_{dd} = \Lambda_{dd} \left( L_z / l_{\perp} \right)$,  
and thus 
$t_{\rm cr}$ is also inversely proportional to the linear number density along $z$. 
This follows from the relation between $\Lambda_{dd} \left( L_z / l_{\perp} \right)$ 
and the linear number density along $z$ displayed in Eq.~\eqref{dipole_interaction_explicit_Lambda}. 

For large dipolar interaction, %that is when $S \Lambda'_{d} \gg \left\vert b_z \right\vert$, 
the asymptotic expression for $t_{\rm cr}$ is, assuming  $\Gamma \ll 1$
\begin{eqnarray}
t_{\rm cr} & \simeq & 
\frac{1}{\Gamma S \Lambda'_{dd}} 
\ln \left[
\frac{5\sqrt{2 (1 - ( M_z)_{\rm in}^2)}
}
{\left\vert \left( M_z \right)_{\rm in} - \left( M_z \right)_{\rm cr} \right\vert}
\right]
%\nonumber 
\label{t_cr_asymp_simple}
\\
& & 
\textrm{provided}\quad S \Lambda'_{dd} \gg \left\vert b_z \right\vert \quad \Longleftrightarrow \quad 
|\left( M_z \right)_{\rm cr}| \ll 1. \nonumber
%\nonumber\\
\end{eqnarray}
The above $t_{\rm cr}$ diverges at $\left( M_z \right)_{\rm in} = \left( M_z \right)_{\rm cr}$ or $\pm 1$, as expected,  since $M_z = \left( M_z \right)_{\rm cr}$ and $M_z = \pm 1$ are time-independent solutions of the LLG equation.
We stress that Eq.~\eqref{t_cr_asymp_simple} clearly shows that the magnetic dipole-dipole interaction {\it accelerates} the decay of $M_z$. Hence, by using a dipolar spinor BEC with large magnetic dipole moment such as produced from ${}^{164} \textrm{Dy}$ or ${}^{166} \textrm{Er}$\, one may observe the relaxation of $M_z$ to the stable state within the BEC lifetime, 
enabling the measurement of $\Gamma$.
% during the lifetime of the BEC. 

Before we show how the critical switching time 
$t_{\rm cr}$ depends on $\left( M_z \right)_{\rm in}$ and $\Gamma$, we will qualitatively discuss when our quasi-1D assumption and 
homogeneous-local-spin-orientation 
assumption are valid. 
Typically, spin-spin-interaction couplings are much smaller than their density-density-interaction
counterparts, by two orders of magnitude. For spin 1 ${}^{23} \textrm{Na}$ BEC or spin 1 ${}^{87} \textrm{Rb}$ BEC, $c_{0} \simeq 100 \left\vert c_{2} \right\vert$~\cite{Uedareview,Palacios_2018}. Thus we may neglect to a first approximation the 
%$c_{2k}$ 
$S^2$ times $c_{2k}$ terms in Eq.~\eqref{quasi_1d_gp_diss_maintext} (see the discussion at the end of Appendix~\ref{magnetostriction-appendix}). 
We also require $\left\vert \left( M_z \right)_{\rm cr} \right\vert < 1$. Thus, we may additionally neglect the $\boldsymbol{b}$ term compared to the $P_{dd} \left( z, t \right)$ %\qq=? $\Rightarrow$ Added more explanation below:
term 
since, for $\boldsymbol{b} = b_{z} \boldsymbol{e}_{z}$, $S \Lambda'_{dd} > \left\vert \boldsymbol{b} \right\vert$ should be satisfied to make $\left\vert \left( M_z \right)_{\rm cr} \right\vert < 1$ (see Eq.~\eqref{def_M_c}) and $\Lambda'_{dd}$ is related to $P_{dd} \left( z, t \right)$ by Eq.~\eqref{lambdap_def}.
When $\Gamma = 0$, 
% and $\boldsymbol{B} = B_{z} \boldsymbol{e}_{z}$, supposing that $\boldsymbol{M} \left( t \right)$ lies along $z$, by writing $\Psi \left( z, t \right) \zeta_{\alpha} \left( t \right) = \varphi_{\alpha} \left( z, t \right)$, 
using our ansatz in Eq.~\eqref{quasi_1d_wavefunc} and integrating out the $x$ and $y$ directions, 
Eq.~\eqref{GP_SW_nodiss-st2} can be approximated by the expression 
\begin{widetext}
\begin{eqnarray}
\mu \left( t \right) 
\Psi \left( \boldsymbol{z}, t \right) 
= && \; 
\left\{ 
- \frac{\hbar^2}{2 m} \frac{\partial^2}{\partial z^2} 
+ V \left( z \right) 
+ \frac{c_0}{2 \pi l_{\perp}^2} \left\vert \Psi \left( z, t \right) \right\vert^2 
+ \Phi_{dd} \left( z, t \right) 
\right\} 
\Psi \left( z, t \right) ,
\label{quasi_1d_gp_diss_maintextapprox}
\end{eqnarray}
where, from Eqs.~\eqref{Phi-dd-def},~\eqref{V_d_real_eff}, and~\eqref{dipole_interaction_explicit_gen}, the dipole-dipole interaction mean-field potential reads 
\begin{eqnarray}
\Phi_{dd} \left( z, t \right) = && \; 
\hbar S^2 \left\{ 1 - 3 M_z^2 \left( t \right) \right\} P_{dd} \left( z, t \right) 
\nonumber\\
= && \; 
\frac{c_{dd}}{2 l_{\perp}^3} 
S^2 \left\{ 1 - 3 M_z^2 \left( t \right) \right\} 
\int_{- \infty}^{\infty} d z' \; 
\left\vert \Psi \left( z', t \right) \right\vert^2 
\left\{ 
G \left( \frac{\left\vert z' - z \right\vert}{l_{\perp}} \right)
- \frac{4}{3} \delta \left( \frac{z' - z}{l_{\perp}} \right) 
\right\} 
\nonumber\\
= && \; 
\frac{c_{dd}}{2 \pi  l_{\perp}^2} 
\pi S^2 \left\{ 1 - 3 M_z^2 \left( t \right) \right\} 
\left\{ 
\int_{- \infty}^{\infty} d \bar{z} \; 
\left\vert \Psi \left( z + \bar{z} l_{\perp}, t \right) \right\vert^2 
G \left( \left\vert \bar{z} \right\vert \right)
- \frac{4}{3} 
\left\vert \Psi \left( z, t \right) \right\vert^2 
\right\} .
\label{Phi_dd-P_dd-connection}
\end{eqnarray}
\end{widetext}
From Fig.~\ref{G-plot}, the function 
$G \left( \lambda \right)$ is positive and decreases exponentially as $\lambda$ increases. 
Thus, if $l_{\perp}$ is small enough such that $\left\vert \Psi \left( z + \bar{z} l_{\perp}, t \right) \right\vert^2$ does not change within the range $\left\vert \bar{z} \right\vert \le 5$, 
one may conclude that 
\begin{eqnarray}
\Phi_{dd} \left( z, t \right) \simeq && \; 
\frac{2 \pi}{3} 
S^2 \left\{ 1 - 3 M_z^2 \left( t \right) \right\} 
\frac{c_{dd}}{2 \pi  l_{\perp}^2} 
\left\vert \Psi \left( z, t \right) \right\vert^2 , \qquad 
\label{P_approx}
\end{eqnarray}
due to the property $\int_{0}^{\infty} d \lambda \; G \left( \lambda \right) = 1$.

A spinor ($S = 6$) dipolar BEC has been realized using  ${}^{166} \textrm{Er}$~\cite{Er_quasi_1D_exp}. 
For this BEC, $c_0 = 4 \pi \hbar^2 a / m$ where $a \simeq 67 \,a_{B}$ ($a_{B}$ is Bohr radius) and 
$2 \pi S^2 c_{dd} / 3 = 0.4911 \,c_0$. 
Due to $\left\vert M_z \left( t \right) \right\vert \le 1$ from the definition of $\boldsymbol{M} \left( t \right)$, 
%and $-S \le \alpha \le S$, 
the maximum value of the chemical potential 
$\mu \left( t \right)$ 
is achieved when $M_z \left( t \right) = 0$, where 
\begin{equation}
\mu \left( t \right) \simeq 
V \left( z \right) + 
\left( c_{0} + \frac{2 \pi}{3} S^2 c_{dd} \right) 
\frac{n \left( z, t \right)}{2 \pi l_{\perp}^2} .
\label{chem-estimation-quasi1D}
\end{equation} 
From above Eq.~\eqref{chem-estimation-quasi1D}, we may regard the 3D number density as $n \left( z, t \right) / \left( 2 \pi l_{\perp}^2 \right)$. 
In~\cite{Er_quasi_1D_exp}, 
$N = 1.2 \times 10^5$, 
$\omega_{\perp} / \left( 2 \pi \right) = \sqrt{156 \times 198}\textrm{\,Hz}%\qq
= 175.75 \textrm{\,Hz}$, 
$\omega_{z} / \left( 2 \pi \right) = 17.2 \textrm{\,Hz}$, 
$l_{\perp} = 0.589 \,\mu$m, and the measured peak number density $\bar{n}_{\rm peak}$ is $6.2 \times 10^{20} \textrm{\,m}^{-3}$. %\qq made $m$ roman.
Using Eq.~\eqref{quasi_1d_gp_diss_maintextapprox} and~\eqref{P_approx}, 
by denoting $L_z$ as the Thomas-Fermi radius along $z$, 
$\left( -L_z \le z \le L_z \right)$ with $V \left( z \right) = m \omega_z^2 z^2 / 2$, 
one derives 
\begin{eqnarray}
L_z = && \; 
\left\{ 
\frac{3 \left( c_{0} + 2 \pi S^2 c_{dd} / 3 \right) N}{4 \pi m \omega_z^{2} l_{\perp}^2} 
\right\}^{1/3} ,
\label{Lz_TF}
\end{eqnarray}
and the mean number density $\bar{n} = \left( N / 2 L_z \right) / \left( 2 \pi l_{\perp}^2 \right) = 6.721 \times 10^{20} \textrm{\,m}^{-3}$ 
as well as chemical potential $\mu / \left( \hbar \omega_{\perp} \right) = m \omega_z^2 L_z^2 / \left( 2 \hbar \omega_{\perp} \right) = 23.22$. 
Note that 
$\bar{n} \simeq 1.1 \,\bar{n}_{\rm peak}$. 
% which implies that our rough calculation is valid. 
Because $\mu$ is not less than $\hbar \omega_{\perp}$, the experiment 
\cite{Er_quasi_1D_exp} is not conducted within the quasi-1D limit.

The 
homogeneous-local-spin-orientation 
approximation is valid when the system size is 
on the order of the spin healing length $\xi_{s}$ or less, which has been experimentally verified in in~\cite{Palacios_2018}. 
Using $c_{0} \simeq 100 \,\left\vert c_{2} \right\vert$, $\xi_{s} \simeq 10 \,\xi_{d}$ where $\xi_{d} = \sqrt{\hbar^2 / \left( 2 m c_0 \bar{n} \right)}$ is the density healing length and $\xi_{s} = \sqrt{\hbar^2 / \left( 2 m \left\vert c_{2} \right\vert \bar{n} \right)}$ is the spin healing length. 
Thus, if $L_z$ is %\qq about
on the order of $10 \,\xi_{d}$, 
the 
homogeneous-local-spin-orientation 
approximation is justified.

Using the $S = 6$ element ${}^{166} \textrm{Er}$, 
we can provide numerical values which satisfy both the quasi-1D and 
homogeneous-local-spin-orientation 
limits,
 as well as they enable us to explicitly show 
 how $t_{\rm cr}$ depends on $\left( M_z \right)_{\rm in}$ in a concretely realizable setup. 
We consider below two cases: (A) box trap along $z$ \footnote{We note, while box traps so far
have been created for scalar BECs with  contact interaction only, 
there is no obstacle in principle to create them as well 
for dipolar spinor gases (J. Dalibard, private communication).} 
and (B) harmonic trap along $z$. 

\subsection{\label{Box-trap-consideration}Box traps}
We set $V \left( z \right) = 0$ for $\left\vert z \right\vert < L_z$ and $\infty$ otherwise. Then $n \left( z, t \right) = N / \left( 2 L_z \right)$ and we estimate 
$\mu \simeq \left( c_{0} + 2 \pi S^2 c_{dd} / 3 \right) N / \left( 4 \pi l_{\perp}^2 L_z \right)$ 
from Eq.~\eqref{chem-estimation-quasi1D}. 
In this case, $\Lambda'_{dd} = \Lambda_{dd} \left( L_z / l_{\perp} \right)$ as is calculated in Eq.~\eqref{dipole_interaction_explicit_Lambda}. 
Fixing $B_z = -0.03\,$mG and $N = 100$, we consider the following two cases: 
(1) $\omega_{\perp} / \left( 2 \pi \right) = 2.4 \times 10^4\,$Hz and $L_z = 3.125 \,\mu$m. 
Then $L_z / l_{\perp} = 62.03$, 
$\mu / \left( \hbar \omega_{\perp} \right) = 0.1692$, 
and $L_z / \xi_{d} = 29.55$. 
Thus, the system is in both the quasi-1D and 
homogeneous-local-spin-orientation 
limit. 
$S \Lambda_{dd} \left( L_z / l_{\perp} \right) = 4.074 \times 10^3\,$Hz, $\hbar S \Lambda_{dd} \left( L_z / l_{\perp} \right) / \left( g_{F} \mu_{B} \right) = 0.3969\,$mG, and $\theta_{\rm cr} \coloneqq \cos^{-1} \left( M_z \right)_{\rm cr}$ is $85.67\,^{\circ}$.

\begin{figure} [t]
\centering
\subfigure{\includegraphics[width=0.4\textwidth]{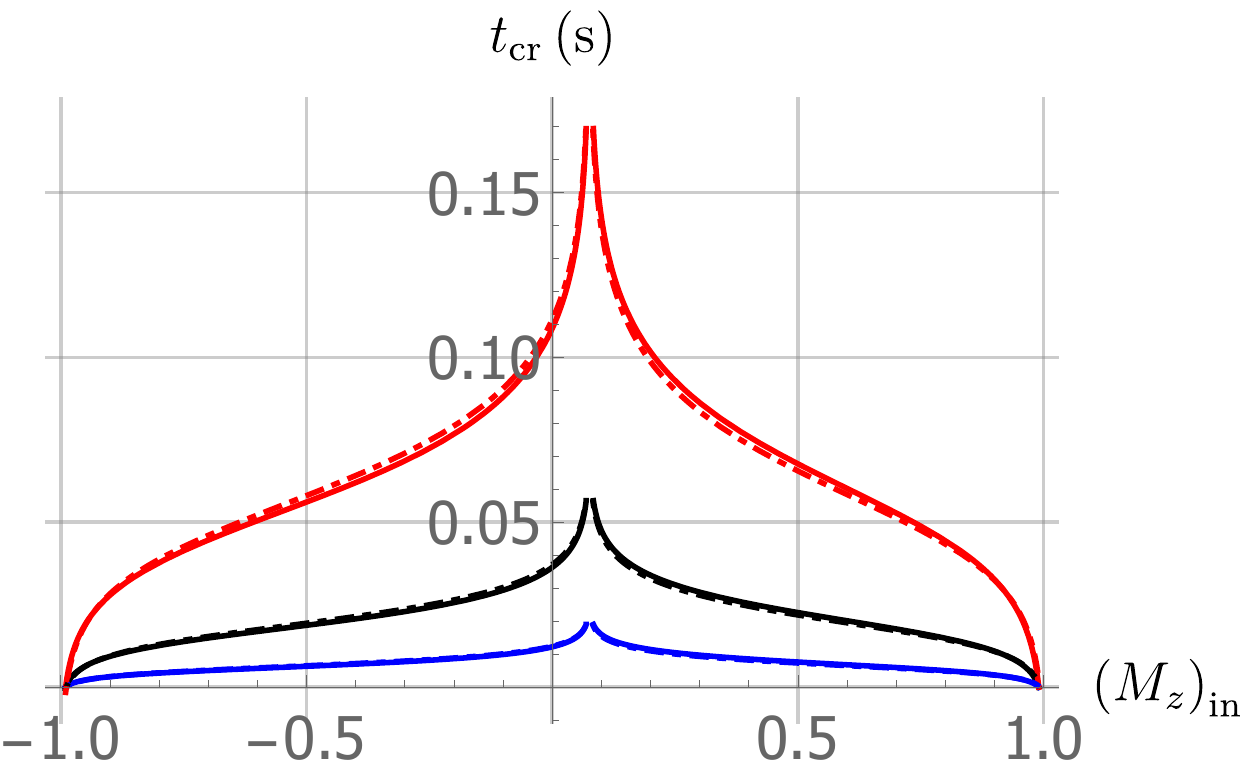}}
{$\omega_{\perp} / \left( 2 \pi \right) = 2.4 \times 10^4\,$Hz, $L_z = 3.125 \,\mu$m, 
and $l_{\perp} = 0.0504 \,\mu$m where $N / \left( 4 \pi L_z l_{\perp}^2 \right) = 10.03 \times 10^{20} \textrm{\,m}^{-3}$ ($\left( M_z \right)_{\rm cr} = 0.0756$).}
\subfigure{\includegraphics[width=0.4\textwidth]{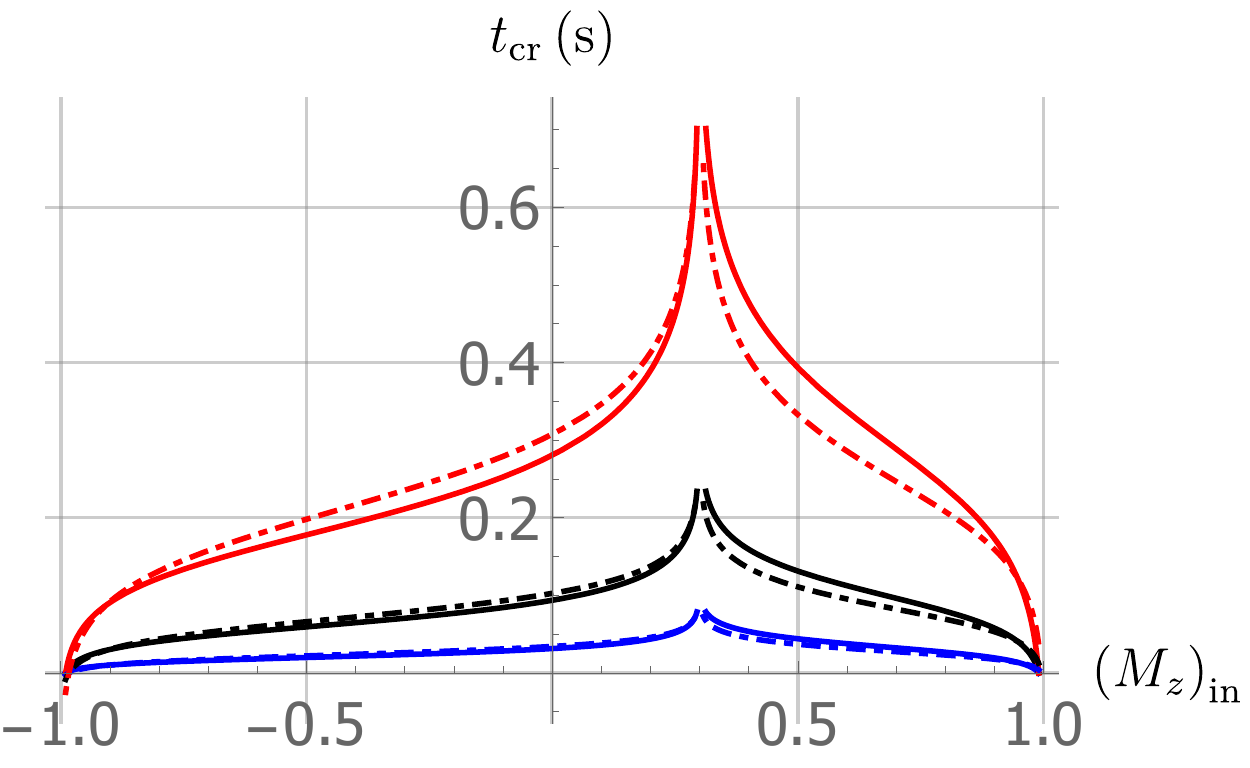}}
{$\omega_{\perp} / \left( 2 \pi \right) = 1.2 \times 10^4\,$Hz, $L_z = 6.250 \,\mu$m, 
and $l_{\perp} = 0.0712 \,\mu$m where $N / \left( 4 \pi L_z l_{\perp}^2 \right) = 2.508 \times 10^{20} \textrm{\,m}^{-3}$ 
($\left( M_z \right)_{\rm cr} = 0.2995$).}
\caption{\label{tcr_Min_graph-box} $t_{\rm cr}$ as a function of $\left( M_z \right)_{\rm in}$ when $\boldsymbol{B} = B_z \boldsymbol{e}_z$ where $B_z = -0.03\,$mG and particle number $N = 100$. 
From top to bottom: 
Red for $\Gamma = 0.01$, black for $\Gamma = 0.03$, and blue for $\Gamma = 0.09$. 
Lines are from {\it exact analytic} formula in Eq.~\eqref{Mz_analytic_sol}, and dot-dashed are from asymptotic expression in Eq.~\eqref{t_cr_asymp_simple}.
Generally, $t_{\rm cr}$ decreases as $\Gamma$ increases. Also, note that $t_{\rm cr}$ diverges as $\left( M_z \right)_{\rm in} \rightarrow \left( M_z \right)_{\rm cr}$. 
For larger mean number density 
$N / \left( 4 \pi L_z l_{\perp}^2 \right)$ 
(top), the asymptotic expression of $t_{\rm cr}$ is 
essentially indistinguishable from the {\it exact analytic} formula of $t_{\rm cr}$.}
\end{figure}

(2) $\omega_{\perp} / \left( 2 \pi \right) = 1.2 \times 10^4\,$Hz and $L_z = 6.250 \,\mu$m. 
Then $L_z / l_{\perp} = 87.72$, 
$\mu / \left( \hbar \omega_{\perp} \right) = 0.0846$, 
and $L_z / \xi_{d} = 29.55$. 
Thus, again the system is in both the quasi-1D and homogeneous-local-spin-orientation  
limits. 
$S \Lambda_{dd} \left( L_z / l_{\perp} \right) = 1.028 \times 10^3\,$Hz, $\hbar S \Lambda_{dd} \left( L_z / l_{\perp} \right) / \left( g_{F} \mu_{B} \right) = 0.1002\,$mG, and $\theta_{\rm cr} = 72.57\,^{\circ}$.
Fig.~\ref{tcr_Min_graph-box} shows the relation between $t_{\rm cr}$ and $\left( M_z \right)_{\rm in}$.

\subsection{Harmonic traps}
We set $V \left( z \right) = m \omega_z^2 z^2 / 2$. Using the Thomas-Fermi approximation, 
from Eq.~\eqref{chem-estimation-quasi1D}, $\mu = m \omega_z^2 L_z^2 / 2$ where $L_z$ is given by Eq.~\eqref{Lz_TF}. $\left( c_{0} + 2 \pi S^2 c_{dd} / 3 \right) n \left( z, t \right) / \left( \pi l_{\perp}^2 \right) = m \omega_z^2 \left( L_z^2 - z^2 \right)$ 
for $\left\vert z \right\vert \le L_z$ and $n \left( z, t \right) = 0$ for $\left\vert z \right\vert > L_z$. 
From this $n \left( z, t \right)$, 
we performed a numerical integration to calculate $\Lambda'_{dd}$ in Eq.~\eqref{lambdap_def}. Fixing $B_z = -0.03\,$mG, we consider the following two cases: 

(1) $N=240$, $\omega_{\perp} / \left( 2 \pi \right) = 2000\,$Hz, and $\omega_z / \left( 2 \pi \right) = 50\,$Hz, for which 
$L_z = 5.703 \,\mu$m and $L_z / l_{\perp} = 32.68$. 
We obtain again the quasi-1D and 
homogeneous-local-spin-orientation  
limits since 
$\mu / \left( \hbar \omega_{\perp} \right) = 0.3337$ and $L_z / \xi_{d} = 17.85$. 
Furthermore, $S \Lambda'_{dd} = 1.644 \times 10^3\,$Hz, $\hbar S \Lambda'_{dd} / \left( g_{F} \mu_{B} \right) = 1.602 \times 10^{-1}\,$mG, and $\theta_{\rm cr} = 79.21\,^{\circ}$. 

(2) $N = 340$, $\omega_{\perp} / \left( 2 \pi \right) = 1000\,$Hz, and $\omega_z / \left( 2 \pi \right) = 25\,$Hz, where 
%=$L_z = 8.070 \mu$m 
$L_z = 8.070 \,\mu$m 
and $L_z / l_{\perp} = 32.70$. 
Again, we have the quasi-1D and 
with homogeneous-local-spin-orientation  limits fulfilled due to 
$\mu / \left( \hbar \omega_{\perp} \right) = 0.3341$ and $L_z / \xi_{d} = 17.87$. 
In addition, $S \Lambda'_{dd} = 8.230 \times 10^2\,$Hz, $\hbar S \Lambda'_{dd} / \left( g_{F} \mu_{B} \right) = 8.019 \times 10^{-2}\,$mG, and $\theta_{\rm cr} = 68.03\,^{\circ}$.

\begin{figure} [t]
\centering
\subfigure{%\label{tcr_Min_graph_N_0_1}
\includegraphics[width=0.4\textwidth]{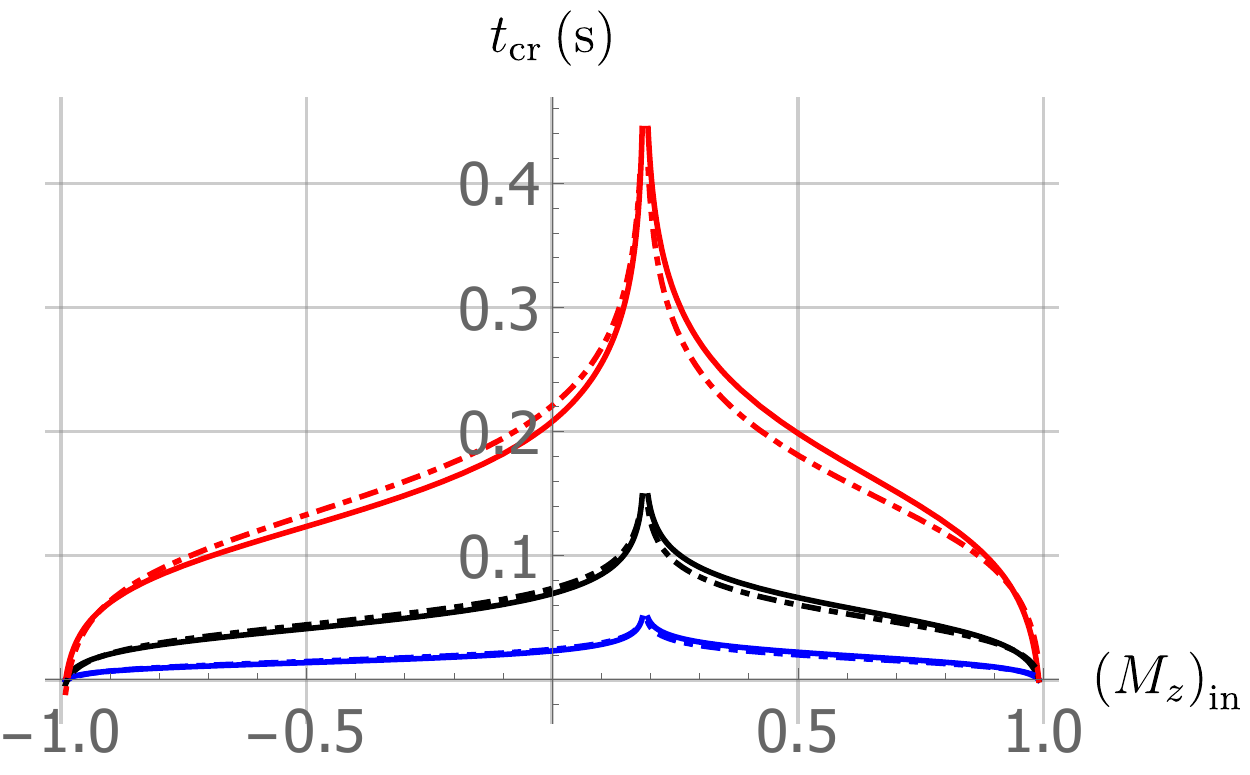}}
{\quad $N=240$, $\omega_{\perp} / \left( 2 \pi \right) = 2000\,$Hz, and $\omega_z / \left( 2 \pi \right) = 50\,$Hz.
$L_z = 5.703 \,\mu$m and $l_{\perp} = 0.1745 \,\mu$m where $N / \left( 4 \pi L_z l_{\perp}^2 \right) = 1.010 \times 10^{20} \textrm{\,m}^{-3}$ ($\left( M_z \right)_{\rm cr} = 0.1873$).
}
\subfigure{%\label{tcr_Min_graph}
\includegraphics[width=0.4\textwidth]{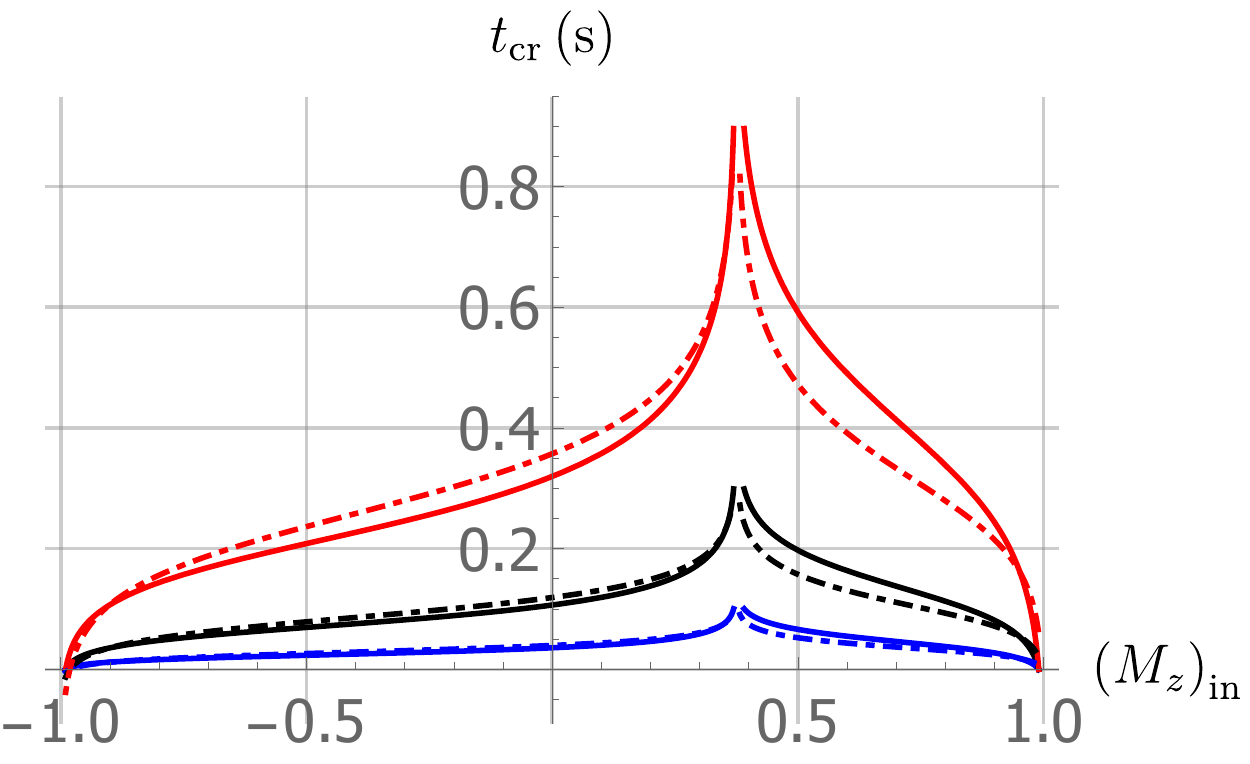}}
{\quad 
$N = 340$, $\omega_{\perp} / \left( 2 \pi \right) = 1000\,$Hz, and $\omega_z / \left( 2 \pi \right) = 25\,$Hz. 
$L_z = 8.070 \,\mu$m and $l_{\perp} = 0.2468 \,\mu$m where $N / \left( 4 \pi L_z l_{\perp}^2 \right) = 0.550 \times 10^{20} \textrm{\,m}^{-3}$ ($\left( M_z \right)_{\rm cr} = 0.3741$).
}
\caption{\label{tcr_Min_graph} $t_{\rm cr}$ as a function of $\left( M_z \right)_{\rm in}$ when $\boldsymbol{B} = B_z \boldsymbol{e}_z$ where $B_z = -0.03\,$mG, for two particle numbers $N$ as shown. 
%where $B_z = -1$ mG, for two particle numbers $N$ as shown. 
From top to bottom: 
Red for $\Gamma = 0.01$, black for $\Gamma = 0.03$, and blue for $\Gamma = 0.09$. 
Lines are from {\it exact analytic} formula in Eq.~\eqref{Mz_analytic_sol}, and dot-dashed are from asymptotic expression in Eq.~\eqref{t_cr_asymp_simple}.
Generally, $t_{\rm cr}$ decreases as $\Gamma$ increases. Also, note that $t_{\rm cr}$ diverges as $\left( M_z \right)_{\rm in} \rightarrow \left( M_z \right)_{\rm cr}$. 
For larger mean number density 
$N / \left( 4 \pi L_z l_{\perp}^2 \right)$ 
(top), the asymptotic expression of $t_{\rm cr}$ is 
essentially indistinguishable from the {\it exact analytic} formula of $t_{\rm cr}$.}
\end{figure}

Fig.~\ref{tcr_Min_graph} shows for the harmonic traps 
the relation between $t_{\rm cr}$ and $\left( M_z \right)_{\rm in}$.

\subsection{Measurability of critical switching time} 
Figs.~\ref{tcr_Min_graph-box} and~\ref{tcr_Min_graph} %and~\ref{tcr_Min_graph_N_0_1} 
demonstrate that
the critical switching time $t_{\rm cr}$ is much smaller than the lifetime of BEC 
(several seconds~\cite{Er_quasi_1D_exp}) and thus, 
by measuring $t_{\rm cr}$ by varying $\left( M_z \right)_{\rm in}$, 
one will be able to obtain the value of $\Gamma$,
 provided $\Gamma$ indeed does not depend on spin indices as for example 
Refs.~\cite{JiHQV,PhysRevA.84.043607} have assumed. Conversely, if one
obtains from the measurements a different functional relation which does not follow 
Eq.~\eqref{Mz_analytic_sol}, this implies that $\Gamma$ may depend on spin indices.

Note that both figures, Figs.~\ref{tcr_Min_graph-box} and~\ref{tcr_Min_graph}, 
show that $t_{\rm cr}$ is {\it inversely} proportional to the mean number density 
$N / \left( 4 \pi L_z l_{\perp}^2 \right)$ . 
Eq.~\eqref{t_cr_asymp_simple} states that $t_{\rm cr}$ is inversely proportional to $\Lambda'_{dd}$, but except for the box trap case, in which one can {\it analytically} calculate $\Lambda'_{dd} = \Lambda_{dd} \left( L_z / l_{\perp} \right)$ in Eq.~\eqref{dipole_interaction_explicit_Lambda}, the dependence of $\Lambda'_{dd}$ and the mean number density 
$N / \left( 4 \pi L_z l_{\perp}^2 \right)$ 
is not immediately apparent.  
Thus, at least for harmonic traps, and in the Thomas-Fermi approximation, one may use the box trap results of Eq.~\eqref{dipole_interaction_explicit_Lambda} 
for provide an approximate estimate of the behavior of $t_{\rm cr}$. 

%We note that $t_{\rm cr}$ is of the order of 10 ms in Figs.~\ref{tcr_Min_graph}, that is  
%and~\ref{tcr_Min_graph_N_0_1}, when the number of condensed atoms of the central core $N_{\rm core}$ is approximately constant~\cite{Er_quasi_1D_exp}. The equations~\eqref{Mz_analytic_sol} and~\eqref{t_cr_asymp_simple} are obtained assuming that number density in $z$ axis $n \left( z, t \right)$ is also constant in time. When the effective magnetic dipole-dipole interaction is not large enough such that $1 / \Gamma S \Lambda'_{dd}$ is, say, of order 100 ms, one may not assume a constant in time linear number density along $z$. % axis $n \left( z, t \right)$. A numerical calculation to solve the LLG equation is then warranted. 

%Due to $N_{\rm core}$ being almost constant within 10 ms, during that time interval, we may neglect the magnetization from noncondensed atoms.

\section{\label{concl}Conclusion}
For a quasi-1D dipolar spinor condensate with unidirectional local magnetization 
(that is in the homogeneous-local-spin-orientation  
limit), we provided an analytical derivation of the Landau-Lifshitz-Gilbert 
equation and the Stoner-Wohlfarth 
model. For an 
external magnetic field along the long axis, we obtained an exact 
solution of the  quasi-1D 
Landau-Lifshitz-Gilbert 
equation.
%which {\it do not change its form whether BEC is confined in quasi-1D or not}. 
Our analytical solution demonstrates that the magnetic dipole-dipole interaction {\it accelerates} the relaxation of the magnetization to stable states and hence strongly facilitates %\qq the possibilty to observe
observation of this process within the lifetime of typical dipolar spinor BECs.
Employing this solution, we hence propose a method to experimentally 
access the dissipative parameter(s) $\Gamma$. 

%So far, we could not find any theoretical basis why $\Gamma \simeq 0.03$ is independent of %spin indices~\cite{JiHQV,PhysRevA.84.043607}. However, that value is from scalar ${}%^{23}$Na BEC~\cite{PhysRevA.57.4057} and hence $\Gamma$ may be differ for spinor BEC.
We expect, in particular, that our proposal provides a viable tool to verify in experiment whether $\Gamma$ is {\it indeed} independent of spin indices, as commonly assumed,  
and does not have to be replaced by a tensorial quantity for spinor gases. 
We %\qq expect
hope that this %\qq would
will stimulate further more detailed investigations %\qq on
of the dissipative mechanism in dipolar BECs with internal degrees of freedom.

We considered that the magnetization along $z$, $M_z$, 
has contributions solely from the atoms residing in the condensate, 
an approximation valid at sufficiently low temperatures. 
%, most of the atoms reside in the condensate~\cite{pethick,Uedareview}. 
When the magnetization from noncondensed atoms is not negligible, as 
considered by Ref.~\cite{Zaremba1999} for a contact interacting scalar BEC, correlation  
terms mixing the noncondensed part and the mean field, 
such as $\textstyle {\sum_{\beta = -S}^{S} \psi^{*}_{\beta} \left( \boldsymbol{r}, t \right) \langle \delta \hat{\psi}_{\alpha} \left( \boldsymbol{r}, t \right) \delta \hat{\psi}_{\beta} \left( \boldsymbol{r}, t \right) \rangle}$ will appear on the right-hand side of Eq.~\eqref{GP_SW_q}. Here, $\delta \hat{\psi}_{\alpha} \left( \boldsymbol{r}, t \right)$ is 
the $\alpha$-th component of  quantum field  excitations above the mean-field ground state in the spinor basis. 
Considering the effect of these terms is a subject of future studies.

\begin{acknowledgments}
The work of SHS 
was supported by the National Research Foundation of Korea (NRF), Grant No. NRF-2015-033908 (Global PhD Fellowship Program). SHS also acknowledges 
the hospitality of the University of T\"ubingen during his stay in the summer of 2019. 
URF has been supported by the NRF under Grant No.~2017R1A2A2A05001422
and Grant No.~2020R1A2C2008103. 
\end{acknowledgments}

\begin{widetext}
\appendix

\section{\label{G_deriv}Derivation of the effective potential $V_{\rm eff}$}
%\bigskip
%By following the definition of effective interaction introduced by~\cite{PhysRevA.84.063633}
%please describe in more detail how this definition proceeds
%{$\leftarrow$ It is not needed to cite~\cite{PhysRevA.84.063633} here since Eq.~\eqref{V_d_real_eff} is written in~\cite{Uedareview} and we will start from that.}
%\bigskip
The dipole-dipole interaction term $V_{dd} \left( t \right) $ in the total energy
is given by~\cite{Uedareview} 
%\qq I find the use of d\boldsymbol{r} for a volume element misleading. Why not write d^3r etc. 
\begin{eqnarray}
V_{dd} \left( t \right) = && \; \frac{c_{dd}}{2} 
%\int d \boldsymbol{r} \int d \boldsymbol{r'} 
\int d^3 r
\int d^3 r'  
\sum_{\nu, \nu' = x, y, z}  
F_{\nu} \left( \boldsymbol{r}, t \right) 
Q_{\nu, \nu'} \left( \boldsymbol{r} - \boldsymbol{r'} \right) 
F_{\nu'} \left( \boldsymbol{r'}, t \right), 
\label{V_d_real_eff}
\end{eqnarray}
where $c_{dd}$ is dipole-dipole interaction coefficient, $F_{\nu} \left( \boldsymbol{r}, t \right) = \psi^{\dagger} \left( \boldsymbol{r}, t \right) \hat{f}_{\nu} \psi \left( \boldsymbol{r}, t \right)$,  
and $Q_{\nu, \nu'} \left( \boldsymbol{r} \right)$ is defined as the tensor %~\cite{Uedareview}
\begin{eqnarray}
Q_{\nu, \nu'} \left( \boldsymbol{r} \right) \coloneqq && \; \frac{r^2 \delta_{\nu, \nu'} - 3 r_{\nu} r_{\nu'}}{r^5}
\label{def_Q_nu_nu}
\end{eqnarray}
in spin space, where $r = \left\vert \boldsymbol{r} \right\vert$ and $r_{\nu} = \boldsymbol{r} \cdot \boldsymbol{e}_{\nu}$, with $\boldsymbol{e}_{\nu}$ being the unit vector along the $\nu$ axis.  
From now on, we %will
define $\boldsymbol{\rho} = \left( x, y \right)$ such that $dx dy = 
%d \boldsymbol{\rho} 
d^2 \rho 
= d \varphi d \rho \; \rho$ where $\tan \varphi = y / x$.

Using the convolution theorem, the dipole-dipole interaction term $V_{dd} \left( t \right)$ can be 
expressed by 
\begin{eqnarray}
V_{dd} \left( t \right) = 
\frac{c_{dd}}{2} 
\left( 2 \pi \right)^{D/2} 
%\int d \boldsymbol{k} \; 
\int d^3 k \; 
\tilde{n} \left( \boldsymbol{k}, t \right) 
\tilde{n} \left( -\boldsymbol{k}, t \right) 
\tilde{U}_{dd} \left( \boldsymbol{k}, t \right) 
\qquad 
\label{V_d_Fourier_conv}
\end{eqnarray}
with the Fourier  transform 
\begin{eqnarray}
U_{dd} \left( \boldsymbol{\eta}, t \right) = && \; 
\frac{1}{n \left( \boldsymbol{r}, t \right) n \left( \boldsymbol{r'}, t \right)} 
\sum_{\nu, \nu' = x, y, z} 
F_{\nu} \left( \boldsymbol{r}, t \right) 
Q_{\nu, \nu'} \left( \boldsymbol{\eta} \right) 
F_{\nu'} \left( \boldsymbol{r'}, t \right), 
\label{U_dd_realspace-def}
\end{eqnarray}
where $\tilde{g} \left( \boldsymbol{k}, t \right) = \left( 2 \pi \right)^{-D/2} \int d \boldsymbol{r} \; g \left( \boldsymbol{r}, t \right) e^{i \boldsymbol{k} \cdot \boldsymbol{r}}$ is the Fourier transform of the function $g \left( \boldsymbol{r}, t \right)$ in $D$-dimensional space $\boldsymbol{r}$ (in our case, $D = 3$), $\boldsymbol{\eta} = \boldsymbol{r} - \boldsymbol{r'}$, 
and $n \left( \boldsymbol{r}, t \right) = \left\vert \psi \left( \boldsymbol{r}, t \right) \right\vert^2$.

By denoting 
$\boldsymbol{k} = \left( \boldsymbol{k}_{\rho}, k_z \right)$, where $\boldsymbol{k}_{\rho} = \left( k_x, k_y \right)$ with $k_{\rho} = \sqrt{k_x^2 + k_y^2}$ and $\tan \varphi_{k_{\rho}} = k_y / k_x$, with our mean-field wavefunction in Eq.~\eqref{quasi_1d_wavefunc}, one derives 
\begin{eqnarray}
\tilde{n} \left( \boldsymbol{k}, t \right) = && \; 
\frac{1}{\pi l_{\perp}^2} 
\frac{1}{\left( 2 \pi \right)^{3/2}} 
%\int d \boldsymbol{\rho} 
\int d^2 \rho 
\int_{-\infty}^{\infty} d z \; 
e^{- \left( \rho / l_{\perp} \right)^2} 
n \left( z, t \right) 
e^{i \boldsymbol{\rho} \cdot \boldsymbol{k}_{\rho}} 
e^{i k_z z}
=
\frac{1}{2 \pi} 
\tilde{n} \left( k_z, t \right) 
e^{- k_{\rho}^2 l_{\perp}^2 / 4},
\label{nkz_fourier}
\end{eqnarray}
where $n \left( z, t \right) \coloneqq \left\vert \Psi \left( z, t \right) \right\vert^2 e^{- 2 \Gamma \omega_{\perp} t / \left( 1 + \Gamma^2 \right)}$. 
Note the  factor of $(2\pi)^{-1}$  appearing, when compared to Eq.~(12) in Ref.~\cite{Giovanazzi2004}, which is  
stemming from our definition of Fourier transform.

Denoting $\eta = \left\vert \boldsymbol{\eta} \right\vert$, by writing $\boldsymbol{e}_{\boldsymbol{\eta}}$ for the unit vector along $\boldsymbol{\eta}$, we obtain
%$U_{dd} \left( \boldsymbol{\eta}, t \right)$ is written as 
\begin{eqnarray}
U_{dd} \left( \boldsymbol{\eta}, t \right) = && \; 
- \frac{1}{\eta^3} \sqrt{\frac{6 \pi}{5}} 
\left\lbrack 
\left\{ 
Y_{2}^{2} \left( \boldsymbol{e}_{\boldsymbol{\eta}} \right) 
e^{- 2 i \phi \left( t \right)} 
+ Y_{2}^{-2} \left( \boldsymbol{e}_{\boldsymbol{\eta}} \right) 
e^{2 i \phi \left( t \right)} 
\right\} 
S^2 \sin^2 \theta \left( t \right)\right. \nonumber\\
&&
\qquad \qquad \qquad - 
\left.\left\{ 
Y_{2}^{1} \left( \boldsymbol{e}_{\boldsymbol{\eta}} \right) 
e^{- i \phi \left( t \right)} 
- Y_{2}^{-1} \left( \boldsymbol{e}_{\boldsymbol{\eta}} \right) 
e^{i \phi \left( t \right)} 
\right\} 
S^2 \sin \left\{ 2 \theta \left( t \right) \right\} 
\right\rbrack 
\nonumber\\
&& \quad + \frac{1}{\eta^3} \sqrt{\frac{6 \pi}{5}} 
Y_{2}^{0} \left( \boldsymbol{e}_{\boldsymbol{\eta}} \right) 
\sqrt{\frac{2}{3}} 
S^2 \left\{ 
3 \sin^2 \theta \left( t \right) - 2 
\right\} ,
\label{U_dd_realspace-form}
\end{eqnarray}
where $Y_{l}^{m} \left( \boldsymbol{e}_{\boldsymbol{\eta}} \right)$ are the usual spherical harmonics. 
Its Fourier transform $\tilde{U}_{dd} \left( \boldsymbol{k}, t \right)$ is 
\begin{eqnarray}
\tilde{U}_{dd} \left( \boldsymbol{k}, t \right) = && \; 
\frac{1}{\left( 2 \pi \right)^{3/2}} \frac{4 \pi}{3} 
S^2 \left\{ 
1 - \frac{3}{2} \sin^2 \theta \left( t \right) 
\right\} 
\left( 3 \frac{k_z^2}{k_{\rho}^2 + k_z^2} - 1 \right) 
\nonumber\\
&& + \frac{1}{\sqrt{2 \pi}} 
\frac{k_{\rho}^2}{k_{\rho}^2 + k_z^2} 
S^2 \sin^2 \theta \left( t \right) 
\cos \left\{ 2 \varphi_{k_{\rho}} - 2 \phi \left( t \right) \right\} 
+ 
\sqrt{\frac{2}{\pi}} 
\frac{k_{\rho} k_z}{k_{\rho}^2 + k_z^2} 
S^2 \sin \left\{ 2 \theta \left( t \right) \right\} 
\cos \left\{ \varphi_{k_{\rho}} - \phi \left( t \right) \right\} .
\label{U_dd_Fourier}
\end{eqnarray}
By plugging Eq.~\eqref{nkz_fourier} and Eq.~\eqref{U_dd_Fourier} into Eq.~\eqref{V_d_Fourier_conv}, we 
finally obtain $V_{dd} \left( t \right)$ as 
\begin{eqnarray}
V_{dd} \left( t \right) = && \; 
\frac{c_{dd}}{2} 
\sqrt{2 \pi} 
\int_{-\infty}^{\infty} d k_z \; 
\tilde{n} \left( k_z, t \right) 
\tilde{n} \left( - k_z, t \right) 
\frac{2 S^2}{l_{\perp}^2 \sqrt{2 \pi}} 
\left\{ 1 - \frac{3}{2} \sin^2 \theta \left( t \right) \right\} 
\left\{
\left( k_z^2 l_{\perp}^2 / 2 \right) 
e^{k_z^2 l_{\perp}^2 / 2} 
E_{1} \left( k_z^2 l_{\perp}^2 / 2 \right) 
- \frac{1}{3} 
\right\}, \quad\quad
\label{V_d_Fourier_conv_calc}
\end{eqnarray}
where $E_1 \left( x \right) = \int_{x}^{\infty} d u \; e^{-u} / u$ is exponential integral.

Note that Eq.~\eqref{V_d_Fourier_conv_calc} can be also written as 
\begin{eqnarray}
V_{dd} \left( t \right) = && \; \frac{c_{dd}}{2} \sqrt{2 \pi} \int_{-\infty}^{\infty} d k_z \; \tilde{n} \left( k_z, t \right) \tilde{n} \left( - k_z, t \right) \tilde{V}_{\rm eff} \left( k_z, t \right) 
= \frac{c_{dd}}{2} 
\int_{- \infty}^{\infty} d z \int_{- \infty}^{\infty} d z' \; 
n \left( z, t \right) n \left( z', t \right) V_{\rm eff} \left( z - z', t \right). 
\label{Veff_Vd_connec}
\end{eqnarray} 
Due to the fact that $\tilde{V}_{\rm eff} \left( k_z, t \right)$ can be obtained by Eq.~\eqref{V_d_Fourier_conv_calc}, we can get $V_{\rm eff} \left( z, t \right)$ by inverse Fourier transform. As a preliminary step, we first write down some 
integrals of $E_1 \left( x \right)$ as follows: 
\begin{eqnarray}
\int_{-\infty}^{\infty} d x \; e^{x^2} E_1 \left( x^2 \right) e^{- i k x} 
= && \;
\int_{- \infty}^{\infty} d x \; e^{- i k x} \int_{x^2}^{\infty} d t \; \frac{e^{- \left( t - x^2 \right)}}{t} 
= \left( \pi \right)^{3/2} e^{k^2 / 4} \textrm{Erfc} \left( \left\vert k \right\vert / 2 \right). 
\label{E1_fourier_prem}
\end{eqnarray}
Differentiating Eq.~\eqref{E1_fourier_prem} with respect to $k$ two times results in
\begin{eqnarray}
\int_{-\infty}^{\infty} d x \; x^2 e^{x^2} E_1 \left( x^2 \right) e^{- i k x} = && \; 
- \left( \pi \right)^{3/2} 
\left\{
\frac{1}{2} \left( \frac{k^2}{2} + 1 \right) e^{k^2 / 4} \textrm{Erfc} \left( \left\vert k \right\vert / 2 \right) 
- \frac{\left\vert k \right\vert}{2 \sqrt{\pi}} 
- \frac{2}{\sqrt{\pi}} \delta \left( k \right) 
\right\}.
\end{eqnarray}
Therefore, $V_{\rm eff} \left( z, t \right)$ can be calculated as
\begin{eqnarray}
V_{\rm eff} \left( z, t \right) 
= && \; 
\frac{1}{\sqrt{2 \pi}} \int_{-\infty}^{\infty} d k_z \; 
\frac{2 S^2}{l_{\perp}^2 \sqrt{2 \pi}} 
\left\{ 1 - \frac{3}{2} \sin^2 \theta \left( t \right) \right\} 
\left\{
\left( k_z^2 l_{\perp}^2 / 2 \right) 
e^{k_z^2 l_{\perp}^2 / 2} 
E_{1} \left( k_z^2 l_{\perp}^2 / 2 \right) 
- \frac{1}{3} 
\right\} e^{- i k_z z}
\nonumber\\
= && \; 
\frac{S^2}{l_{\perp}^3} 
\left\{ \frac{3}{2} \sin^2 \theta \left( t \right) - 1 \right\} 
\left\{
G \left( \left\vert z \right\vert / l_{\perp} \right) - \frac{4}{3} \delta \left( z / l_{\perp} \right) 
\right\},
\label{V_eff_real_space}
\end{eqnarray}
where $G \left( x \right)$ is defined in Eq.~\eqref{F_defin}, and $\delta \left( x \right)$ is the Dirac delta function.

%We will calculate the  Fourier transform of Eq.~\eqref{V_eff_real_space} to see 
%whether it is correct result.
The Fourier transform %$\tilde{V}_{\rm eff} \left( k_z, t \right)$ 
of Eq.~\eqref{V_eff_real_space} acquires the form  
\begin{eqnarray}
\tilde{V}_{\rm eff} \left( k_z, t \right) = && \; 
\frac{1}{\sqrt{2 \pi}} 
\int_{- \infty}^{\infty} d z \; 
V_{\rm eff} \left( z, t \right) e^{i k_z z} 
= 
\sqrt{\frac{2}{\pi}} 
\frac{S^2}{l_{\perp}^2} 
\left\{ \frac{3}{2} \sin^2 \theta \left( t \right) - 1 \right\} 
\left\{
\int_{0}^{\infty} d v \; G \left( v \right) \cos \left( k_z l_{\perp} v \right) 
- \frac{2}{3}
\right\}
\nonumber\\
= && \; 
\sqrt{\frac{2}{\pi}} 
\frac{S^2}{l_{\perp}^2} 
\left\{ \frac{3}{2} \sin^2 \theta \left( t \right) - 1 \right\} 
\left\lbrack
\int_{0}^{\infty} d u \; 
\left\{
\sqrt{\pi} \left( 2 u^2 + 1 \right) e^{u^2} \textrm{Erfc} \left( u \right) - 2 u
\right\}
 \cos \left( \sqrt{2} k_z l_{\perp} u \right) 
- \frac{2}{3}
\right\rbrack.
\label{eff_V_Fourier_1}
\end{eqnarray} 
From~\cite{geller1969table}, the following integral involving the complementary error function is 
\begin{eqnarray}
\int_{0}^{\infty} d u \; e^{u^2} \textrm{Erfc} \left( u \right) \cos \left( b u \right) = \frac{1}{2 \sqrt{\pi}} e^{b^2 / 4} E_1 \left( b^2 / 4 \right).
\label{erfc_cos_integ}
\end{eqnarray}
By differentiating Eq.~\eqref{erfc_cos_integ} two times with respect to $b$, we get
\begin{eqnarray}
\int_{0}^{\infty} d u \; u^2 e^{u^2} \textrm{Erfc} \left( u \right) \cos \left( b u \right) = 
- \frac{1}{2 \sqrt{\pi}} \left\{
\frac{1}{2} \left( \frac{b^2}{2} + 1 \right) e^{b^2 / 4} E_1 \left( b^2 / 4 \right) - 1 + \frac{2}{b^2}
\right\}.
\end{eqnarray}
Hence, Eq.~\eqref{eff_V_Fourier_1} becomes 
\begin{eqnarray}
\tilde{V}_{\rm eff} \left( k_z, t \right) 
= && \; 
\sqrt{\frac{2}{\pi}} 
\frac{S^2}{l_{\perp}^2} 
\left\{ \frac{3}{2} \sin^2 \theta \left( t \right) - 1 \right\} 
\left\lbrack
- \left\{
\frac{1}{2} \left( k_z^2 l_{\perp}^2 + 1 \right) e^{k_z^2 l_{\perp}^2 / 2} E_1 \left( k_z^2 l_{\perp}^2 / 2 \right) - 1 + \frac{1}{k_z^2 l_{\perp}^2}
\right\} \right. \nonumber\\
& & 
+ \left.\frac{1}{2} e^{k_z^2 l_{\perp}^2 / 2} E_1 \left( k_z^2 l_{\perp}^2 / 2 \right) 
+ \frac{1}{k_z^2 l_{\perp}^2} - \frac{2}{3}
\right\rbrack
\nonumber\\
= && \; 
\frac{2 S^2}{l_{\perp}^2 \sqrt{2 \pi}} 
\left\{ 1 - \frac{3}{2} \sin^2 \theta \left( t \right) \right\}  
\left\{
\left( k_z^2 l_{\perp}^2 / 2 \right) e^{k_z^2 l_{\perp}^2 / 2} E_1 \left( k_z^2 l_{\perp}^2 / 2 \right) 
- \frac{1}{3}
\right\}.
\label{eff_V_Fourier_res}
\end{eqnarray} 
Comparing Eq.~\eqref{V_d_Fourier_conv_calc} with Eq.~\eqref{eff_V_Fourier_res}, one verifies  that Eq.~\eqref{V_eff_real_space} is the correct result for the effective interaction of the quasi-1D dipolar spinor gas.

\section{\label{quasi_1d_llg_deriv}Quasi-1D Gross-Pitaevski\v\i~equation with dissipation}
By introducing an identical damping coefficient for each component of the spinor, cf., 
e.g.~Refs.~\cite{JiHQV,PhysRevA.84.043607} (i.e.~as if each component effectively 
behaves as a scalar BEC~\cite{PhysRevA.67.033610}), and neglecting a possible %\qq when there is no 
quadratic Zeeman term, the GP equation for a spin-$S$ BEC can be written as~\cite{PhysRevA.84.043607}
\begin{eqnarray}
\left( i - \Gamma \right) \hbar \frac{\partial \psi_\alpha \left( \boldsymbol{r}, t \right)}{\partial t} = && \; 
\left\{ 
- \frac{\hbar^2}{2 m} \nabla^2 
+ V_{\rm tr} \left( \boldsymbol{r} \right) 
+ c_0 \left\vert \psi \left( \boldsymbol{r}, t \right) \right\vert^2 
\right\} 
\psi_\alpha \left( \boldsymbol{r}, t \right) 
- \hbar \sum_{\beta = -S}^{S} 
\left\{
\boldsymbol{b} 
- \boldsymbol{b}_{dd} \left( \boldsymbol{r}, t \right) 
\right\} 
\cdot 
\left( \boldsymbol{\hat{f}} \right)_{\alpha, \beta} \psi_{\beta} \left( \boldsymbol{r}, t \right)
\nonumber\\
&& \quad + 
\sum_{k = 1}^{S} c_{2 k} 
\sum_{\nu_1, \nu_2, \cdots, \nu_k = x, y, z} 
F_{\nu_1, \nu_2, \cdots, \nu_k} \left( \boldsymbol{r}, t \right) 
\sum_{\beta = -S}^{S} 
\left( \hat{f}_{\nu_1} \hat{f}_{\nu_2} \cdots \hat{f}_{\nu_k} \right)_{\alpha,\beta} 
\psi_{\beta} \left( \boldsymbol{r}, t \right),
\nonumber\\
\label{GP_SW}
\end{eqnarray}
where $\psi_\alpha \left( \boldsymbol{r}, t \right)$ is the $\alpha$-th component of the mean-field wavefunction $\psi \left( \boldsymbol{r}, t \right)$ (the spin-space index $\alpha$ is an integer taking $2S+1$ values running from $-S$ and $S$), 
$F_{\nu_1, \nu_2, \cdots, \nu_k} \left( \boldsymbol{r}, t \right) \coloneqq 
\psi^{\dagger} \left( \boldsymbol{r}, t \right) 
\hat{f}_{\nu_1} \hat{f}_{\nu_2} \cdots \hat{f}_{\nu_k} 
\psi \left( \boldsymbol{r}, t \right)$, 
$\hbar \boldsymbol{\hat{f}}$ is the spin-$S$ operator, 
$\boldsymbol{b} = g_F \mu_B \boldsymbol{B} / \hbar$ ($g_F$ is the Land\'e g-factor, 
$\mu_B$ is the Bohr magneton, and $\boldsymbol{B}$ the external magnetic field).
Finally, 
$\hbar \boldsymbol{b}_{dd} \left( \boldsymbol{r}, t \right) \cdot \boldsymbol{e}_{\nu} = 
c_{dd} 
%\int d \boldsymbol{r'} \; 
\int d^3 r' \; 
\sum_{\nu' = x, y, z} 
Q_{\nu, \nu'} \left( \boldsymbol{r} - \boldsymbol{r'} \right) 
F_{\nu'} \left( \boldsymbol{r'}, t \right)
$, where $\boldsymbol{e}_{\nu}$ is the unit vector along the $\nu$ axis ($\nu = x, y, z$)~\cite{Uedareview}. 
Applying the formalism of Ref.~\cite{pitaevskii1959phenomenological} to a spinor BEC assuming that $\Gamma$ does not depend on spin indices, one just needs to transform $t \rightarrow \left( 1 + \Gamma^2 \right) t$ in Eq.~\eqref{GP_SW} and~\eqref{quasi_1d_wavefunc}. We then  integrate out the $x$ and $y$ directions in Eq.~\eqref{GP_SW} to obtain the 
quasi-1D GP equation.

From Eq.~\eqref{quasi_1d_wavefunc} in the main text, we have 
\begin{multline}
%\int d \boldsymbol{\rho} 
\int d^2 \rho 
\sum_{\beta = -S}^{S}  
\frac{e^{- \rho^2 / \left( 2 l_{\perp}^2 \right)}}{l_{\perp} \sqrt{\pi}}  
\left\{ 
\hbar \boldsymbol{b}_{dd} \left( \boldsymbol{r} \right) \cdot \left( \boldsymbol{\hat{f}} \right)_{\alpha, \beta} \right\} \psi_{\beta} \left( \boldsymbol{r}, t \right) \\
= \frac{c_{dd}}{2 l_{\perp}^3} 
\int_{- \infty}^{\infty} d z' \; 
n \left( z', t \right) 
%\\\times 
\left\{
G \left( \frac{\left\vert z - z' \right\vert}{l_{\perp}} \right) - \frac{4}{3} \delta \left( \frac{z - z'}{l_{\perp}} \right) 
\right\} 
\Psi \left( z, t \right) e^{- \frac{i + \Gamma}{1 + \Gamma^2} \omega_{\perp} t} 
S
\left\{ 
\boldsymbol{M} \left( t \right) - 3 M_z \left( t \right) \boldsymbol{e}_{z}  
\right\} \cdot \sum_{\beta = -S}^{S} \left( \boldsymbol{\hat{f}} \right)_{\alpha, \beta} \zeta_{\beta} \left( t \right) ,
\end{multline}
where 
$\int d^2 \rho \coloneqq \int_{- \infty}^{\infty} d x \int_{- \infty}^{\infty} d y$ and 
$n \left( z, t \right) \coloneqq 
%\int d \boldsymbol{\rho} \;  
\int d^2 \rho \;
\left\vert \psi \left( \boldsymbol{r}, t \right) \right\vert^2 
= \left\vert \Psi \left( z, t \right) \right\vert^2 e^{- 2 \Gamma \omega_{\perp} t / \left( 1 + \Gamma^2 \right)}$. 

For a spin-$S$ BEC, from Eq.~\eqref{GP_SW}, 
for the trap potential given in Eq.~\eqref{trap} and if we use Eq.~\eqref{quasi_1d_wavefunc}, 
by integrating out the $x$ and $y$ directions, one acquires the expression 
\begin{eqnarray}
\left( i - \Gamma \right) \hbar \frac{\partial \left\{ \Psi \left( z, t \right) \zeta_{\alpha} \left( t \right) \right\}}{\partial t} = && \; 
\left\{
- \frac{\hbar^2}{2 m} \frac{\partial^2}{\partial z^2} + V \left( z \right) 
+ \frac{c_0}{2 \pi l_{\perp}^2} n \left( z, t \right)  
\right\} \Psi \left( z, t \right) \zeta_{\alpha} \left( t \right) 
\nonumber\\
&& + \left\lbrack
- \hbar \boldsymbol{b} 
+ \hbar S \left\{ \boldsymbol{M} \left( t \right) - 3 M_z \left( t \right) \boldsymbol{e}_z \right\}  P_{dd} \left( z, t \right) 
\right\rbrack 
\cdot \left\{ 
\sum_{\beta = -S}^{S} \left( \boldsymbol{\hat{f}} \right)_{\alpha, \beta} \Psi \left( z, t \right) \zeta_{\beta} \left( t \right) \right\}
\nonumber\\
&& + \sum_{k = 1}^{S} \frac{c_{2k}}{2 \pi l_{\perp}^2} n \left( z, t \right) \sum_{\nu_1, \nu_2, \cdots, \nu_k = x, y, z} 
S M_{\nu_1, \nu_2, \cdots, \nu_k} \left( t \right) 
\left\{
\sum_{\beta = -S}^{S} \left( \hat{f}_{\nu_1} \hat{f}_{\nu_2} \cdots \hat{f}_{\nu_k} \right)_{\alpha, \beta} 
\Psi \left( z, t \right) \zeta_{\beta} \left( t \right) 
\right\} ,
\nonumber\\
\label{quasi_1d_gp_diss}
\end{eqnarray} 
where $M_{\nu_1, \nu_2, \cdots, \nu_k} \left( t \right)$ is defined in Eq.~\eqref{multiple_Mk} 
and 
\begin{eqnarray}
P_{dd} \left( z, t \right) = && \; 
\frac{c_{dd}}{2 \hbar  l_{\perp}^3} 
\int_{- \infty}^{\infty} d z' \; 
n \left( z', t \right) \left\{ G \left( \frac{\left\vert z - z' \right\vert}{l_{\perp}} \right) - \frac{4}{3} \delta \left( \frac{z - z'}{l_{\perp}} \right) \right\} 
= \frac{c_{dd}}{\hbar S^2 \left\{ 3 \sin^2 \theta \left( t \right) - 2 \right\}} 
\int_{- \infty}^{\infty} d z' \; 
n \left( z', t \right) V_{\rm eff} \left( z - z', t \right) , 
\nonumber\\
\label{Q_def_2}
\end{eqnarray}
with  $V_{\rm eff}$ defined in \eqref{V_eff_real_space}. 
%From Eq.~\eqref{quasi_1d_wavefunc}, $\int d \boldsymbol{\rho} \; \left\vert \psi \left( \boldsymbol{r}, t \right) \right\vert^2 = \left\vert \Psi \left( z, t \right) \right\vert^2 e^{- 2 \Gamma \omega_{\perp} t / \left( 1 + \Gamma^2 \right)}$. Hence, if we renormalize $\psi \left( \boldsymbol{r}, t \right)$ every time so that $\int d \boldsymbol{r} \; \left\vert \psi \left( \boldsymbol{r}, t \right) \right\vert^2 \coloneqq N$ becomes constant in time, $\left\vert \Psi \left( z, t \right) \right\vert^2 e^{- 2 \Gamma \omega_{\perp} t / \left( 1 + \Gamma^2 \right)}$ should be constant in time, which we will call $n \left( z \right)$. Hence, $P \left( z, t \right)$ becomes constant in time (which we will denote as $P \left( z \right)$) when we renormalize $\psi \left( \boldsymbol{r}, t \right)$ every time.
It is already clear from Eq.~\eqref{quasi_1d_gp_diss} that, besides particle loss from the condensate encoded in a decaying $|\Psi(z,t)|$, dissipation also leads to a {\it dephasing}, i.e.~the decay of $\zeta(t)$ due to the term $-\Gamma \partial \zeta(t)/\partial t$.  

From now on, if there is no ambiguity, and for brevity, we drop the arguments such as 
$x,y,z,t$ from the functions.
%simplicity, we %will
%write $g \left( \alpha_1, \alpha_2, ... \right)$ as $g$ where $g \left( \alpha_1, \alpha_2, ... \right)$ is some function $g$ of $\alpha_1$, $\alpha_2$,...,  
%where the $\alpha_i$ 
%stand for some arbitrary variables; e.g., time $t$, position variables $\boldsymbol{r}$, $z$, etc.
%stand for time $t$ and position variables $\boldsymbol{r}$, $z$. 
From Eq.~\eqref{quasi_1d_gp_diss}, we then get 
\begin{eqnarray}
\hbar \frac{\partial \zeta_{\alpha}}{\partial t} = && \; 
- \frac{\hbar}{\Psi} \frac{\partial \Psi}{\partial t} \zeta_{\alpha} 
- \frac{\Gamma + i}{1 + \Gamma^2} 
\left( 
- \frac{\hbar^2}{2 m} \frac{1}{\Psi} \frac{\partial^2 \Psi}{\partial z^2} + V
+ \frac{c_{0}}{2 \pi l_{\perp}^2} n 
\right) \zeta_{\alpha} 
+ \frac{\Gamma + i}{1 + \Gamma^2} \left\{
\hbar \boldsymbol{b} - S \left( \boldsymbol{M} - 3 M_z \boldsymbol{e}_{z} \right) \hbar P_{dd} 
\right\} \cdot \left\{ \sum_{\beta = -S}^{S} \left( \boldsymbol{\hat{f}} \right)_{\alpha, \beta} \zeta_{\beta} \right\} 
\nonumber\\
&& \quad - \frac{\Gamma + i}{1 + \Gamma^2} \sum_{k = 1}^{S} 
\frac{c_{2 k}}{2 \pi l_{\perp}^2} n \sum_{\nu_1, \nu_2, \cdots, \nu_k = x, y, z} S M_{\nu_1, \nu_2, \cdots, \nu_k} 
\left\{
\sum_{\beta = -S}^{S}
\left( \hat{f}_{\nu_1} \hat{f}_{\nu_2} \cdots \hat{f}_{\nu_k} \right)_{\alpha, \beta}
\zeta_{\beta} 
\right\} ,
\label{mean_GP_diss_zeta}
\end{eqnarray}
%where $c_{j}' \coloneqq c_{j} / \left( 2 \pi l_{\perp}^2 \right)$.
Since $\frac{\partial \left\vert \zeta \right\vert^2}{\partial t} = 0$ due to the normalization $\left\vert \zeta \right\vert^2 = 1$, we then have 
\begin{eqnarray}
0 = && \; 
2 \textrm{Re} \left\{ 
- \frac{\hbar}{\Psi} \frac{\partial \Psi}{\partial t} 
- \frac{\Gamma}{1 + \Gamma^2} 
\left( 
- \frac{\hbar^2}{2 m} \frac{1}{\Psi} \frac{\partial^2 \Psi}{\partial z^2} + V 
+ \frac{c_{0}}{2 \pi l_{\perp}^2} n 
\right) 
\right\}
+ \frac{i}{1 + \Gamma^2} \frac{\hbar^2}{2 m} \left( \frac{1}{\Psi} \frac{\partial^2 \Psi}{\partial z^2} - \frac{1}{\Psi^{*}} \frac{\partial^2 \Psi^{*}}{\partial z^2} \right) 
\nonumber\\
&& + \frac{2 \Gamma}{1 + \Gamma^2} \left\{
\hbar \boldsymbol{b} - S \left( \boldsymbol{M} - 3 M_z \boldsymbol{e}_{z} \right) \hbar P_{dd} 
\right\} \cdot S \boldsymbol{M} 
- \frac{2 \Gamma}{1 + \Gamma^2} \sum_{k = 1}^{S} 
\frac{c_{2 k}}{2 \pi l_{\perp}^2} n \sum_{\nu_1, \nu_2, \cdots, \nu_k = x, y, z} S^2 M^2_{\nu_1, \nu_2, \cdots, \nu_k} .
\end{eqnarray}
Hence the dynamics of the magnetization direction follows  the equation 
\begin{eqnarray}
\hbar S \frac{\partial M_{\nu}}{\partial t} 
= && \; 
2 \textrm{Re} \left\{ \sum_{\alpha, \beta = -S}^{S} \zeta_{\alpha}^{\dagger} \left( \hat{f}_{\nu} \right)_{\alpha, \beta} \left( \hbar \frac{\partial \zeta_{\beta}}{\partial t} \right) \right\} 
\nonumber\\
= && \; - \frac{2 \Gamma}{1 + \Gamma^2} S^2 M_{\nu} 
\left\{
\hbar \boldsymbol{b} - S \left( \boldsymbol{M} - 3 M_z \boldsymbol{e}_{z} \right) \hbar P_{dd} 
\right\} \cdot \boldsymbol{M} 
+ \frac{2 \Gamma}{1 + \Gamma^2} M_{\nu} \sum_{k = 1}^{S} 
\frac{c_{2 k}}{2 \pi l_{\perp}^2} n \sum_{\nu_1, \nu_2, \cdots, \nu_k = x, y, z} S^3 M^2_{\nu_1, \nu_2, \cdots, \nu_k} 
\nonumber\\
&& + \frac{\Gamma}{1 + \Gamma^2} \sum_{\mu=x,y,z} \left\{ 
\hbar b_{\mu} - S \left( M_{\mu} - 3 M_{z} \delta_{\mu, z} \right) \hbar P_{dd} 
\right\} 
S \left\{ \delta_{\mu, \nu} + \left( 2 S - 1 \right) M_{\mu} M_{\nu} \right\} 
\nonumber\\
&& - \frac{1}{1 + \Gamma^2} \sum_{\mu,\kappa=x,y,z} \left\{ 
\hbar b_{\mu} - S \left( M_{\mu} - 3 M_{z} \delta_{\mu, z} \right) \hbar P_{dd} 
\right\} 
\epsilon_{\nu, \mu, \kappa} S M_{\kappa}  
\nonumber\\
&& - 2 \textrm{Re} 
\left\{ 
\frac{\Gamma + i}{1 + \Gamma^2} 
\sum_{k = 1}^{S} \frac{c_{2 k}}{2 \pi l_{\perp}^2} n
\sum_{\nu_1, \nu_2, \cdots, \nu_k = x, y, z} S M_{\nu_1, \nu_2, \cdots, \nu_k} 
\sum_{\alpha, \beta = -S}^{S}
\zeta_{\alpha}^{\dagger} 
\left( \hat{f}_{\nu} \hat{f}_{\nu_1} \hat{f}_{\nu_2} \cdots \hat{f}_{\nu_k} \right)_{\alpha, \beta}
\zeta_{\beta} 
\right\} ,
\label{M_diss_eq_st1}
\end{eqnarray}
since the scalar product 
$\zeta^{\dagger} \left( \hat{f}_{\alpha} \hat{f}_{\beta} + \hat{f}_{\beta} \hat{f}_{\alpha} \right) \zeta = S \left\{ \delta_{\alpha, \beta} + \left( 2 S - 1 \right) M_{\alpha} M_{\beta} \right\}$~\cite{PhysRevA.84.043607}.

By direct comparison, % with ~\cite{PhysRevA.84.043607}, 
we can identify Eq.~\eqref{our_B21} below as being identical to Eq.~(B21) in~\cite{PhysRevA.84.043607}, the only difference consisting in the definition of $M_{\nu_1, \nu_2, \cdots, \nu_k}$:  We employ a scaled version of $M_{\nu_1, \nu_2, \cdots, \nu_k}$, which is normalized to $S$ in  \cite{PhysRevA.84.043607}. 
From Eq.~\eqref{wf_zeta_def} in the main text, 
\begin{eqnarray}
\sum_{\nu_1, \nu_2, \cdots, \nu_k = x, y, z} M_{\nu_1, \nu_2, \cdots, \nu_k} 
\sum_{\alpha, \beta = -S}^{S} 
\zeta_{\alpha}^{\dagger} 
\left( \hat{f}_{\nu} \hat{f}_{\nu_1} \hat{f}_{\nu_2} \cdots \hat{f}_{\nu_k} \right)_{\alpha, \beta}
\zeta_{\beta} = && \; 
\sum_{\nu_1, \nu_2, \cdots, \nu_k = x, y, z} M^2_{\nu_1, \nu_2, \cdots, \nu_k} 
S^2 M_{\nu},  
\label{our_B21}
\end{eqnarray}
which is real. Therefore, Eq.~\eqref{M_diss_eq_st1} can be written in the following form \begin{eqnarray}
\frac{\partial \boldsymbol{M}}{\partial t} = && \; 
- \frac{\Gamma}{1 + \Gamma^2} 
\boldsymbol{M} \times 
\left\lbrack
\boldsymbol{M} \times 
\left\{ 
\boldsymbol{b} - S \left( \boldsymbol{M} - 3 M_{z} \boldsymbol{e}_{z} \right) P_{dd} 
\right\} 
\right\rbrack
+ \frac{1}{1 + \Gamma^2} 
\boldsymbol{M} \times 
\left\{ 
\boldsymbol{b} - S \left( \boldsymbol{M} - 3 M_{z} \boldsymbol{e}_{z} \right) P_{dd} 
\right\} 
\nonumber\\
= && \; \frac{1}{1 + \Gamma^2} 
\boldsymbol{M} \times 
\left( 
\boldsymbol{b} + 3 S P_{dd} M_{z} \boldsymbol{e}_{z} 
\right) 
- \frac{\Gamma}{1 + \Gamma^2} 
\boldsymbol{M} \times 
\left\lbrack
\boldsymbol{M} \times 
\left( 
\boldsymbol{b} + 3 S P_{dd} M_{z} \boldsymbol{e}_{z} 
\right) 
\right\rbrack 
\nonumber\\
= && \; \boldsymbol{M} \times 
\left( 
\boldsymbol{b} + 3 S P_{dd} M_{z} \boldsymbol{e}_{z} 
\right) 
- \Gamma \boldsymbol{M} \times 
\frac{\partial \boldsymbol{M}}{\partial t} ,
\label{LLG_bef_integ_z}
\end{eqnarray}
since $\boldsymbol{M} \cdot \frac{\partial \boldsymbol{M}}{\partial t} = 0$ holds. 

As $P$ is a function of 
$z$ and $t$,   
but $\boldsymbol{M}$ is independent of $z$ 
[$\boldsymbol{M}$ is the scaled local magnetization and 
our aim is to study a dipolar spinor BEC with unidirectional 
local magnetization (the homogeneous-local-spin-orientation  
limit)], by multiplying with 
$n \left( z, t \right)$ 
both sides of Eq.~\eqref{LLG_bef_integ_z} and integrating along $z$, we finally get the LLG equation 
\begin{eqnarray}
\frac{\partial \boldsymbol{M}}{\partial t} = && \; \boldsymbol{M} \times 
\left( 
\boldsymbol{b} + S \Lambda'_{dd} M_{z} \boldsymbol{e}_{z} 
\right) 
- \Gamma \boldsymbol{M} \times 
\frac{\partial \boldsymbol{M}}{\partial t},
\label{M-LLG-eq}
\end{eqnarray}
where $\Lambda'_{dd}$ is defined in Eq.~\eqref{lambdap_def}.
Note here that $\Lambda'_{dd}$ becomes $\Lambda_{dd} \left( L_z / l_{\perp} \right)$ 
defined in Eq.~\eqref{dipole_interaction_explicit_Lambda} when 
$n \left( z, t \right) = N / \left( 2 L_z \right)$ for $- L_z \le z \le L_z$ and $n \left( z, t \right) = 0$ otherwise.

\section{\label{Gamma_tensor}Modification of the LLG equation for $\Gamma$ a spin-space tensor}
When $\Gamma$ depends on spin indices, i.e. is a tensor,  
Eq.~\eqref{quasi_1d_gp_diss} can be generalized to read  
\begin{eqnarray}
\sum_{\beta=-S}^{S} 
\left( i \delta_{\alpha, \beta} - \Gamma_{\alpha, \beta} \right) \hbar \frac{\partial \left\{ \Psi \left( z, t \right) \zeta_{\beta} \left( t \right) \right\}}{\partial t} 
= && \; \sum_{\beta=-S}^S H_{\alpha, \beta} \Psi \left( z, t \right) \zeta_{\beta} \left( t \right). 
\label{tensor_G_quasi_1d_gp_diss}
\end{eqnarray} 
The spinor part of the wavefunction is normalized to unity, $\left\vert \zeta \right\vert^2 = 1$. Hence, we know that $\frac{\partial \left\vert \zeta \right\vert^2}{\partial t} = 0$. Therefore, from Eq.~\eqref{tensor_G_quasi_1d_gp_diss}, we derive the expression 
\begin{eqnarray}
\sum_{\alpha, \beta=-S}^S \textrm{Re} 
\left\lbrack
- i \zeta^{*}_{\alpha} \Gamma_{\alpha, \beta} \frac{\partial \zeta_{\beta}}{\partial t} 
- i \zeta^{*}_{\alpha} \Gamma_{\alpha, \beta} \zeta_{\beta} \frac{1}{\Psi} \frac{\partial \Psi}{\partial t} - i \frac{1}{\hbar \Psi} \zeta^{*}_{\alpha} H_{\alpha, \beta} \Psi \zeta_{\beta} 
\right\rbrack  - \textrm{Re} 
\left\lbrack
\frac{1}{\Psi} \frac{\partial \Psi}{\partial t} \right\rbrack
= 0.
\label{tensor_G_deriv_1}
\end{eqnarray} 
This then leads us to 
\begin{eqnarray}
\frac{\partial M_{\nu}}{\partial t} & = & \frac{2}{S} \sum_{\alpha, \beta, \gamma=-S}^S \textrm{Re} 
\left\lbrack 
- i \zeta^{*}_{\alpha} \left( \hat{f}_{\nu} \right)_{\alpha, \beta} \Gamma_{\beta, \gamma} \frac{\partial \zeta_{\gamma}}{\partial t} 
-  i \zeta^{*}_{\alpha} \left( \hat{f}_{\nu} \right)_{\alpha, \beta} \Gamma_{\beta, \gamma} \zeta_{\gamma} \frac{1}{\Psi} \frac{\partial \Psi}{\partial t} 
- i \frac{1}{\hbar \Psi} \zeta^{*}_{\alpha} \left( \hat{f}_{\nu} \right)_{\alpha, \beta} H_{\beta, \gamma} \Psi \zeta_{\gamma} 
\right\rbrack  \nonumber \\ 
& & - 2\textrm{Re} 
\left\lbrack 
M_{\nu}  \frac{1}{\Psi} \frac{\partial \Psi}{\partial t}\right\rbrack  . %\quad\quad 
%\nonumber\\
\label{tensor_G_deriv_2}
\end{eqnarray}
For scalar $\Gamma$, $\Gamma_{\alpha, \beta}\rightarrow \Gamma \delta_{\alpha, \beta}$, the equation above becomes  
%in Eq.~\eqref{tensor_G_deriv_2}. Eq.~\eqref{tensor_G_deriv_2} below 
Eq.~\eqref{M_diss_eq_st1}. %, use $\Gamma_{\alpha, \beta} = \Gamma \delta_{\alpha, \beta}$.

From Eqs.~\eqref{tensor_G_deriv_1} and \eqref{tensor_G_deriv_2}, one concludes that the stationary solution $M_{\nu}$ of Eq.~\eqref{tensor_G_deriv_2} is independent of $\Gamma$. 
In other words, whether $\Gamma$ depends on spin indices or not, the SW model 
\eqref{SW_H} is left unaffected, also see the discussion in Section~\ref{connec_to_SW}
of the main text.

\section{\label{magnetostriction-appendix}
Description of magnetostriction}

For a dipolar spinor BEC without quadratic Zeeman term, when there is no dissipation ($\Gamma = 0$), the mean-field equation in Eq.~\eqref{GP_SW_q} can be written as 
\begin{eqnarray}
\mu_{\alpha} \left( t \right) 
\psi_{\alpha} \left( \boldsymbol{r}, t \right) = && \; 
\left\{ 
- \frac{\hbar^2}{2 m} \nabla^2 
+ V_{\rm tr} \left( \boldsymbol{r} \right) 
+ c_0 \sum_{\beta = -S}^{S} \left\vert \psi_{\beta} \left( \boldsymbol{r}, t \right) \right\vert^2 
\right\} 
\psi_{\alpha} \left( \boldsymbol{r}, t \right) 
- \hbar 
\left\{
\boldsymbol{b} 
- \boldsymbol{b}_{dd} \left( \boldsymbol{r}, t \right) 
\right\} 
\cdot 
\sum_{\beta = -S}^{S} 
\left( \boldsymbol{\hat{f}} \right)_{\alpha, \beta} 
\psi_{\beta} \left( \boldsymbol{r}, t \right)
\nonumber\\
&& \quad + 
\sum_{k = 1}^{S} c_{2 k} 
\sum_{\nu_1, \nu_2, \cdots, \nu_k = x, y, z} 
\sum_{\alpha_1, \beta_1, \beta = -S}^{S} 
\left( \hat{f}_{\nu_1} \hat{f}_{\nu_2} \cdots \hat{f}_{\nu_k} \right)_{\alpha_1, \beta_1} 
\left( \hat{f}_{\nu_1} \hat{f}_{\nu_2} \cdots \hat{f}_{\nu_k} \right)_{\alpha, \beta} 
\psi^{*}_{\alpha_1} \left( \boldsymbol{r}, t \right) 
\psi_{\beta_1} \left( \boldsymbol{r}, t \right) 
\psi_{\beta} \left( \boldsymbol{r}, t \right). 
\nonumber\\
\label{GP_SW_nodiss}
\end{eqnarray}
where we have substituted $i \hbar \frac{\partial \psi_{\alpha} \left( \boldsymbol{r}, t \right)}{\partial t} = \mu_{\alpha} \left( t \right) \psi_{\alpha} \left( \boldsymbol{r}, t \right)$. 

Since we consider the homogeneous-local-spin-orientation  limit, we may write $\psi_{\alpha} \left( \boldsymbol{r}, t \right) = \Psi_{\rm uni} \left( \boldsymbol{r}, t \right) \zeta_{\alpha} \left( t \right)$. 
In this limit, we have 
\begin{eqnarray}
\left\vert \psi \left( \boldsymbol{r}, t \right) \right\vert^2 \coloneqq 
\psi^{\dagger} \left( \boldsymbol{r}, t \right) \psi \left( \boldsymbol{r}, t \right) = \sum_{\alpha = -S}^{S} \psi^{\dagger}_{\alpha} \left( \boldsymbol{r}, t \right) \psi_{\alpha} \left( \boldsymbol{r}, t \right)
= \left\vert \Psi_{\rm uni} \left( \boldsymbol{r}, t \right) \right\vert^2 ,
\end{eqnarray}
since $\displaystyle \sum_{\alpha = -S}^{S} \left\vert \zeta_{\alpha} \left( t \right) \right\vert^2 = 1$ from the definition of $\zeta_{\alpha} \left( t \right)$ in Eq.~\eqref{wf_zeta_def}. 
Thus $\left\vert \Psi_{\rm uni} \left( \boldsymbol{r}, t \right) \right\vert^2$ is equal to the number density.
Then Eq.~\eqref{GP_SW_nodiss} can be written as 
\begin{eqnarray}
\mu_{\alpha} \left( t \right) 
\zeta_{\alpha} \left( t \right) 
\Psi_{\rm uni} \left( \boldsymbol{r}, t \right) 
= && \; 
\left\{ 
- \frac{\hbar^2}{2 m} \nabla^2 
+ V_{\rm tr} \left( \boldsymbol{r} \right) 
+ c_0 \left\vert \Psi_{\rm uni} \left( \boldsymbol{r}, t \right) \right\vert^2 
\right\} 
\zeta_{\alpha} \left( t \right) 
\Psi_{\rm uni} \left( \boldsymbol{r}, t \right) 
\nonumber\\
&& \quad - \hbar 
\left\{
\boldsymbol{b} 
- \boldsymbol{b}_{dd} \left( \boldsymbol{r}, t \right) 
\right\} 
\cdot 
\sum_{\beta = -S}^{S} 
\left( \boldsymbol{\hat{f}} \right)_{\alpha, \beta} 
\zeta_{\beta} \left( t \right) 
\Psi_{\rm uni} \left( \boldsymbol{r}, t \right) 
\nonumber\\
&& \quad + S 
\sum_{k = 1}^{S} c_{2 k} 
\sum_{\nu_1, \nu_2, \cdots, \nu_k = x, y, z} 
\sum_{\beta = -S}^{S} 
M_{\nu_1, \nu_2, \cdots, \nu_k} \left( t \right) 
\left( \hat{f}_{\nu_1} \hat{f}_{\nu_2} \cdots \hat{f}_{\nu_k} \right)_{\alpha, \beta} 
\zeta_{\beta} \left( t \right) 
\left\vert \Psi_{\rm uni} \left( \boldsymbol{r}, t \right) \right\vert^2 
\Psi_{\rm uni} \left( \boldsymbol{r}, t \right) .
\nonumber\\
\label{GP_SW_nodiss-st1}
\end{eqnarray}

Now, we decompose the chemical potential 
$\mu \left( t \right)$ 
as $\displaystyle \mu \left( t \right) \coloneqq \sum_{\alpha = -S}^{S} \mu_{\alpha} \left( t \right) \left\vert \zeta_{\alpha} \left( t \right) \right\vert^2$. 
Then one obtains 
\begin{eqnarray}
\mu \left( t \right) 
\Psi_{\rm uni} \left( \boldsymbol{r}, t \right) 
= && \; 
\left\lbrack 
- \frac{\hbar^2}{2 m} \nabla^2 
+ V_{\rm tr} \left( \boldsymbol{r} \right) 
+ \left\{ 
c_0 
+ S^2 
\sum_{k = 1}^{S} c_{2 k} 
\sum_{\nu_1, \nu_2, \cdots, \nu_k = x, y, z} 
M^2_{\nu_1, \nu_2, \cdots, \nu_k} \left( t \right) 
\right\} 
\left\vert \Psi_{\rm uni} \left( \boldsymbol{r}, t \right) \right\vert^2 
\right\rbrack 
\Psi_{\rm uni} \left( \boldsymbol{r}, t \right) 
\nonumber\\
&& \quad + 
\left\lbrack 
\Phi_{dd} \left( \boldsymbol{r}, t \right) 
- S \hbar 
\left\{ \boldsymbol{b} \cdot \boldsymbol{M} \left( t \right) \right\} 
\right\rbrack 
\Psi_{\rm uni} \left( \boldsymbol{r}, t \right) ,
\nonumber\\
\label{GP_SW_nodiss-st2}
\end{eqnarray}
where 
\begin{eqnarray}
\Phi_{dd} \left( \boldsymbol{r}, t \right) \coloneqq 
S^2 c_{dd} 
\left\lbrack 
%\int d \boldsymbol{r'} 
\int d^3 r' \; 
\left\{ 
\sum_{\nu, \nu' = x, y, z} 
M_{\nu} \left( t \right) 
Q_{\nu, \nu'} \left( \boldsymbol{r} - \boldsymbol{r'} \right) 
M_{\nu'} \left( t \right) 
\right\} 
\left\vert \Psi_{\rm uni} \left( \boldsymbol{r'}, t \right) \right\vert^2 
\right\rbrack ,
\label{Phi-dd-def}
\end{eqnarray}
is the dipole-dipole mean-field potential~\cite{PhysRevLett.89.130401} 
following from the definition of $\boldsymbol{b}_{dd}$ below Eq.~\eqref{GP_SW_q}
in the main text.

Due to $M_{x} \left( t \right) = \sin \theta \left( t \right) \cos \phi \left( t \right)$, $M_{y} \left( t \right) = \sin \theta \left( t \right) \sin \phi \left( t \right)$, and $M_{z} \left( t \right) = \cos \theta \left( t \right)$, from Eqs.~\eqref{U_dd_realspace-def} and~\eqref{U_dd_realspace-form}, we have   
\begin{eqnarray}
\sum_{\nu, \nu' = x, y, z} 
M_{\nu} \left( t \right) 
Q_{\nu, \nu'} \left( \boldsymbol{\eta} \right) 
M_{\nu'} \left( t \right) 
= && \; 
- \frac{1}{\eta^3} \sqrt{\frac{6 \pi}{5}} 
\left\lbrack 
\left\{ 
Y_{2}^{2} \left( \boldsymbol{e}_{\boldsymbol{\eta}} \right) 
e^{- 2 i \phi \left( t \right)} 
+ Y_{2}^{-2} \left( \boldsymbol{e}_{\boldsymbol{\eta}} \right) 
e^{2 i \phi \left( t \right)} 
\right\} 
\sin^2 \theta \left( t \right)\right. \nonumber\\
&&
\qquad \qquad \qquad - 
\left.\left\{ 
Y_{2}^{1} \left( \boldsymbol{e}_{\boldsymbol{\eta}} \right) 
e^{- i \phi \left( t \right)} 
- Y_{2}^{-1} \left( \boldsymbol{e}_{\boldsymbol{\eta}} \right) 
e^{i \phi \left( t \right)} 
\right\} 
\sin \left\{ 2 \theta \left( t \right) \right\} 
\right\rbrack 
\nonumber\\
&& \quad + \frac{1}{\eta^3} \sqrt{\frac{6 \pi}{5}} 
Y_{2}^{0} \left( \boldsymbol{e}_{\boldsymbol{\eta}} \right) 
\sqrt{\frac{2}{3}} 
\left\{ 
3 \sin^2 \theta \left( t \right) - 2 
\right\} ,
\label{MQM-form1}
\end{eqnarray}
where $Y_{l}^{m} \left( \boldsymbol{e}_{\boldsymbol{\eta}} \right)$ are the usual spherical harmonics.

By using Eq.~\eqref{def_Q_nu_nu}, an alternative form of Eq.~\eqref{MQM-form1} can be obtained: 
\begin{eqnarray}
\sum_{\nu, \nu' = x, y, z} 
M_{\nu} \left( t \right) 
Q_{\nu, \nu'} \left( \boldsymbol{\eta} \right) 
M_{\nu'} \left( t \right) 
= && \; 
\sum_{\nu, \nu' = x, y, z} 
\frac{\eta^2 \delta_{\nu, \nu'} - 3 \eta_{\nu} \eta_{\nu'}}{\eta^5}
M_{\nu} \left( t \right) 
M_{\nu'} \left( t \right) 
= \frac{\eta^2 \left\vert \boldsymbol{M} \left( t \right) \right\vert^2 - 3 \left\{ \boldsymbol{\eta} \cdot \boldsymbol{M} \left( t \right) \right\}^2}{\eta^5}
\nonumber\\
= && \; 
\frac{\eta^2 - 3 \left\{ \boldsymbol{\eta} \cdot \boldsymbol{M} \left( t \right) \right\}^2}{\eta^5} .
\label{MQM-form2}
\end{eqnarray}

Thus, $\Phi_{dd} \left( \boldsymbol{r}, t \right)$ can be written as 
\begin{eqnarray}
\Phi_{dd} \left( \boldsymbol{r}, t \right) = && \; 
S^2 c_{dd} 
\left\lbrack 
\int d^3 r' \; 
\frac{\left\vert \boldsymbol{r} - \boldsymbol{r'} \right\vert^2 - 3 \left\{ \left( \boldsymbol{r} - \boldsymbol{r'} \right) \cdot \boldsymbol{M} \left( t \right) \right\}^2}{\left\vert \boldsymbol{r} - \boldsymbol{r'} \right\vert^5}
\left\vert \Psi_{\rm uni} \left( \boldsymbol{r'}, t \right) \right\vert^2 
\right\rbrack 
\nonumber\\
= && \; 
S^2 c_{dd} 
\left\lbrack 
\int d^3 \bar{r}' \; 
\frac{\left\vert \boldsymbol{\bar{r}} - \boldsymbol{\bar{r}'} \right\vert^2 - 3 \left\{ \left( \boldsymbol{\bar{r}} - \boldsymbol{\bar{r}'} \right) \cdot \boldsymbol{M} \left( t \right) \right\}^2}{\left\vert \boldsymbol{\bar{r}} - \boldsymbol{\bar{r}'} \right\vert^5}
\left\vert \Psi_{\rm uni} \left( \boldsymbol{\bar{r}'}, t \right) \right\vert^2 
\right\rbrack 
\nonumber\\
= && \; 
- \frac{3}{2} S^2 c_{dd} \sin^2 \theta \left( t \right) 
\int d^3 \bar{\eta} \; 
\left\vert \Psi_{\rm uni} \left( \boldsymbol{\bar{\eta}} + \boldsymbol{\bar{r}}, t \right) \right\vert^2 
\frac{1}{\bar{\eta}^5} 
\left\lbrack 
\bar{\eta}^2 - \bar{\eta}_z^2 
- 2 
\left\{ 
\bar{\eta}_x \sin \phi \left( t \right) 
- \bar{\eta}_y \cos \phi \left( t \right) 
\right\}^2 
\right\rbrack 
\nonumber\\
&& \quad - 
3 S^2 c_{dd} \sin \left\{ 2 \theta \left( t \right) \right\} 
\int d^3 \bar{\eta} \; 
\left\vert \Psi_{\rm uni} \left( \boldsymbol{\bar{\eta}} + \boldsymbol{\bar{r}}, t \right) \right\vert^2 
\frac{\bar{\eta}_z}{\bar{\eta}^5} 
\left\{ 
\bar{\eta}_x \cos \phi \left( t \right) 
+ \bar{\eta}_y \sin \phi \left( t \right) 
\right\} 
\nonumber\\
&& \quad + 
\frac{1}{2} S^2 c_{dd} 
\left\{ 1 - 3 \cos^2 \theta \left( t \right) \right\} 
\int d^3 \bar{\eta} \; 
\left\vert \Psi_{\rm uni} \left( \boldsymbol{\bar{\eta}} + \boldsymbol{\bar{r}}, t \right) \right\vert^2 
\frac{1}{\bar{\eta}^5} 
\left( 3 \bar{\eta}_z^2 - \bar{\eta}^2 \right) .
\label{Phi_dd_form}
\end{eqnarray}
where $\bar{\boldsymbol{r}} \coloneqq \boldsymbol{r} / L$ with $L$ being some length which scales $r$ (so that $\bar{\boldsymbol{r}}$ is a dimensionless vector). For example, in quasi-1D with trap potential being Eq.~\eqref{trap}, $L = l_{\perp}$. 
Note that, in the special case where $\boldsymbol{M} \left( t \right) = M_z \left( t \right) \boldsymbol{e}_{z}$, the form of Eq.~\eqref{Phi_dd_form} becomes identical to Eq.(6) in Ref.~\cite{sapina_ground-state_2010}. 

Since we concentrate on quasi-1D gases, with trap potential given by Eq.~\eqref{trap} 
in the main text, we will
explicitly compute the form of $\Phi_{dd} \left( \boldsymbol{r}, t \right)$ for the 
quasi-1D setup.
By writing 
\begin{eqnarray}
\left\vert \Psi_{\rm uni} \left( \boldsymbol{r}, t \right) \right\vert^2 = 
\frac{e^{- \rho^2 / l_{\perp}^2}}{\pi l_{\perp}^2} 
\left\vert \Psi \left( z, t \right) \right\vert^2 ,
\end{eqnarray}
and integrating out $x$ and $y$ directions, one can get the quasi-1D dipole-dipole-interaction mean-field potential $\Phi_{dd} \left( z, t \right)$ as follows (which is in Eq.~\eqref{Phi_dd-P_dd-connection}): 
\begin{eqnarray}
\Phi_{dd} \left( z, t \right) = && \; 
\frac{c_{dd}}{2 l_{\perp}^2} 
S^2 \left\{ 1 - 3 M_z^2 \left( t \right) \right\} 
\left\{ 
\int_{- \infty}^{\infty} d \bar{z} \; 
\left\vert \Psi \left( z + \bar{z} l_{\perp}, t \right) \right\vert^2 
G \left( \left\vert \bar{z} \right\vert \right)
- \frac{4}{3} 
\left\vert \Psi \left( z, t \right) \right\vert^2 
\right\} .
\end{eqnarray}

Now, let us consider box trap in quasi-1D case, i.e. $V \left( z \right) = 0$ for $\left\vert z \right\vert \le L_z$ and $V \left( z \right) = \infty$ for $\left\vert z \right\vert > L_z$ where $V \left( z \right)$ is in Eq.~\eqref{trap}. 
Then we may write 
\begin{eqnarray}
\left\vert \Psi \left( z, t \right) \right\vert^2 = 
\left\lbrack 
\begin{array}{cc} 
\displaystyle 
\frac{N}{2 L_z} 
& \textrm{for $\left\vert z \right\vert \le L_z$,} 
\\
\\
0 & \textrm{for $\left\vert z \right\vert > L_z$,}
\end{array}
\right.
\end{eqnarray}
since $V \left( z \right) = 0$ for $- L_z \le z \le L_z$. Thus, $\Phi_{dd} \left( z, t \right)$ can be written as 
\begin{eqnarray}
\Phi_{dd} \left( z, t \right) = && \; 
\left\lbrack 
\begin{array}{cc}
\displaystyle 
\bar{\Phi}_{dd} \left( t \right) 
\left\{ 
\int_{- \left( L_z + z \right) / l_{\perp}}^{\left( L_z - z \right) / l_{\perp}} d \bar{z} \; 
G \left( \left\vert \bar{z} \right\vert \right)
- \frac{4}{3} 
\right\} & \textrm{for $\left\vert z \right\vert \le L_z$,} 
\\
\\
\displaystyle 
\bar{\Phi}_{dd} \left( t \right) 
\int_{- \left( L_z + z \right) / l_{\perp}}^{\left( L_z - z \right) / l_{\perp}} d \bar{z} \; 
G \left( \left\vert \bar{z} \right\vert \right) 
& \textrm{for $\left\vert z \right\vert > L_z$,} 
\end{array}
\right.
\end{eqnarray}
where $\bar{\Phi}_{dd} \left( t \right) \coloneqq N c_{dd} S^2 \left\{ 1 - 3 M_z^2 \left( t \right) \right\} / \left( 2 L_z l_{\perp}^2 \right)$. $\Phi_{dd} \left( z, t \right)$ is discontinuous at $z = \pm L_z$ because of the sudden change of the density at the boundary ($z = \pm L_z$) due to box trap potential. 

\begin{figure} [t]
\centering
\subfigure{
\includegraphics[width=0.4\textwidth]{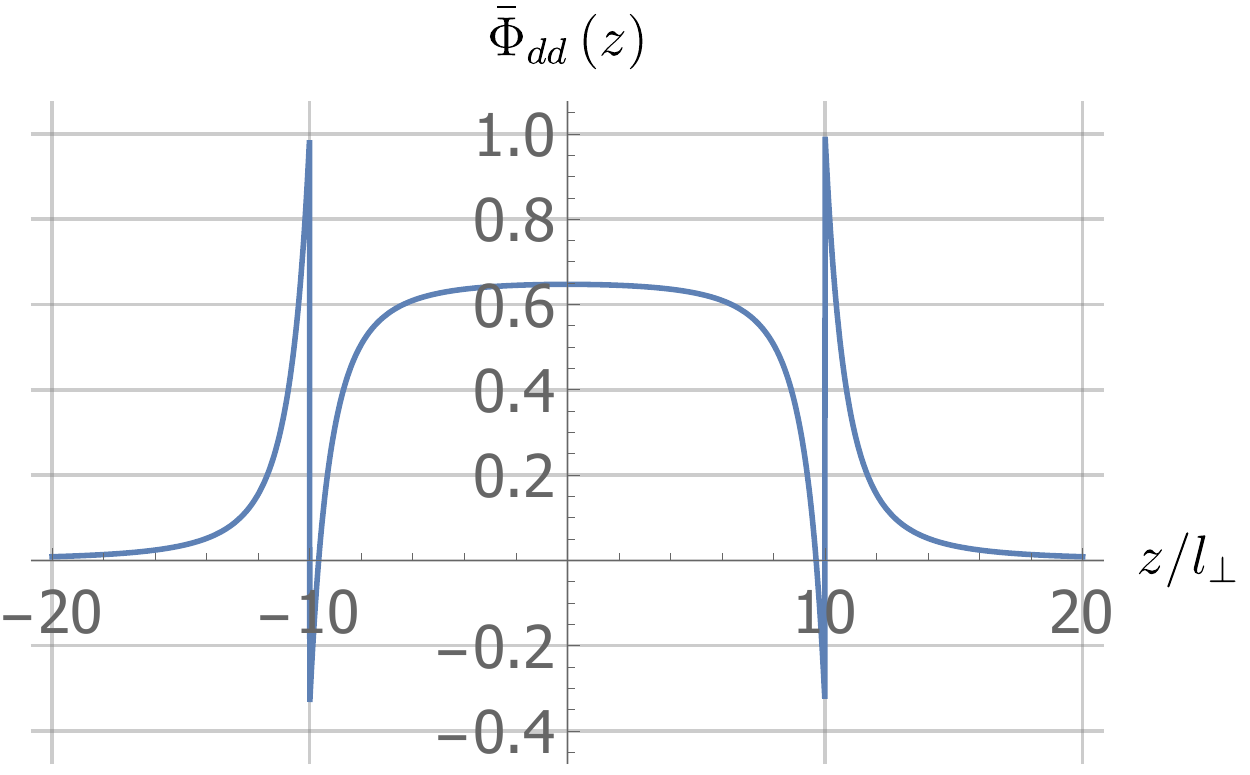}}
\subfigure{
\includegraphics[width=0.4\textwidth]{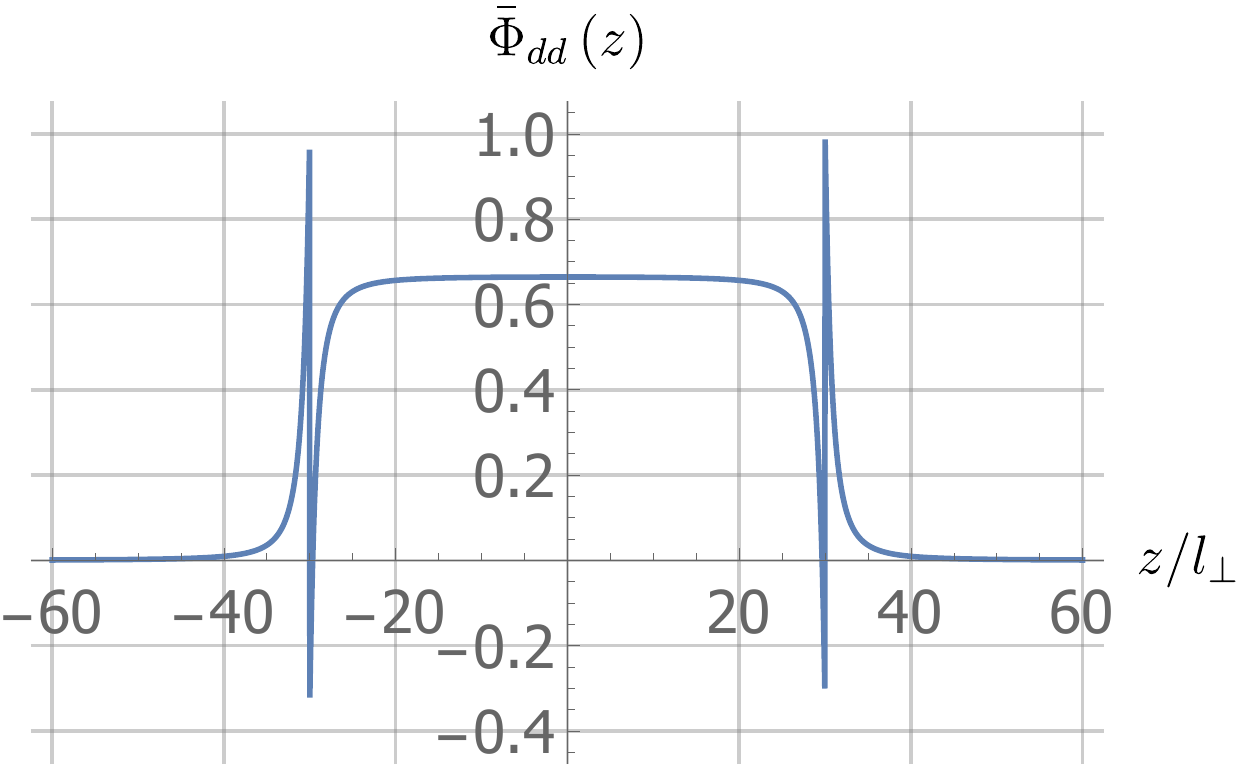}}
\caption{\label{1D-Phidd} Scaled dipole-dipole mean-field potential $\bar{\Phi}_{dd} \left( z \right)$ as a function of $z$ for a quasi-1D box trap. 
(Left) $L_z / l_{\perp} = 10$. (Right) $L_z / l_{\perp} = 30$.}
\end{figure}

Defining the scaled density-density mean-field potential $\bar{\Phi}_{dd} \left( z \right) \coloneqq \Phi_{dd} \left( z, t \right) / \bar{\Phi}_{dd}\left( t \right)$, 
we obtain Fig.~\ref{1D-Phidd}, for two different axial extensions, $L_z / l_{\perp} = 10$ and $30$.  
As Fig.~\ref{1D-Phidd} clearly illustrates, in a box-trapped quasi-1D gas, 
$\Phi_{dd} \left( z, t \right)$ becomes approximately constant for $\left\vert z \right\vert < L_c$ and $L_c \rightarrow L_z$ for $L_z / l_{\perp} \gg 1$. 
Depending on the value of $\boldsymbol{M} \left( t \right)$, $\Phi_{dd} \left( \boldsymbol{r}, t \right)$ will introduce either a repulsive or an attractive force. 
This force will however exist only near the boundary for a box trap, 
%\qq 
where it can lead to a slight modification of the density of atomes.  Its relative influence decreases %\qq the larger the
with increasing extension of the trapped gas along the $z$ axis, and can therefore be consistently neglected in the approximation of constant particle-density. 
%\qq 

However, to assess whether significant magnetostriction occurs, 
%whether the system size will increased or not, 
%as the form of Eq.~\eqref{GP_SW_nodiss-st2} is similar to the mean field equation for scalar BEC, 
one has  to consider, in addition to $\Phi_{dd}$,  
the trap potential $V_{\rm tr}$ % \left( \boldsymbol{r} \right)$ 
and the 
`quasi' density-density interaction mean field potential $\Phi_{0}$ % \left( \boldsymbol{r}, t \right)$ which is 
defined as 
\begin{eqnarray}
\Phi_{0} \left( \boldsymbol{r}, t \right) \coloneqq 
\left\{ 
c_0 
+ S^2 
\sum_{k = 1}^{S} c_{2 k} 
\sum_{\nu_1, \nu_2, \cdots, \nu_k = x, y, z} 
M^2_{\nu_1, \nu_2, \cdots, \nu_k} \left( t \right) 
\right\} 
\left\vert \Psi_{\rm uni} \left( \boldsymbol{r}, t \right) \right\vert^2 .
\end{eqnarray}
We can coin $\Phi_{0} \left( \boldsymbol{r}, t \right)$ a `quasi' density-density interaction mean field potential because only $c_0$ is a density-density interaction coefficient ($c_{2k}$ are interaction coefficients parametrizing the spin-spin interactions for spin-$S$ gas where $k$ is an integer with $1 \le k \le S$. For example, $c_{2}$ is the spin-spin interaction coefficient of a 
spin-1 gas). 
In our quasi-1D case, this $\Phi_{0} \left( \boldsymbol{r}, t \right)$ potential 
is $\Phi_{0} \left( z, t \right)$ where 
\begin{eqnarray}
\Phi_{0} \left( z, t \right) \coloneqq 
\left\{ 
\frac{c_0}{2 \pi l_{\perp}^2} 
+ S^2 
\sum_{k = 1}^{S} \frac{c_{2 k}}{2 \pi l_{\perp}^2} 
\sum_{\nu_1, \nu_2, \cdots, \nu_k = x, y, z} 
M^2_{\nu_1, \nu_2, \cdots, \nu_k} \left( t \right) 
\right\} 
\left\vert \Psi \left( z, t \right) \right\vert^2 .
\end{eqnarray}

%\colb{
In the main text, we assume that $c_{0}\gg \textstyle S^2 \sum_{k = 1}^{S} c_{2 k}\sum_{\nu_1, \nu_2, \cdots, \nu_k = x, y, z} M^2_{\nu_1, \nu_2, \cdots, \nu_k} \left( t \right)$. 
For spin-1 ${}^{23} \textrm{Na}$ or ${}^{87} \textrm{Rb}$, $S = 1$ and $c_{0} \simeq 100 \left\vert c_{2} \right\vert$~\cite{Uedareview,Palacios_2018}, so this is an appropriate assumption (note that $ \sum_{k = 1}^{S} \sum_{\nu_1, \nu_2, \cdots, \nu_k = x, y, z} M^2_{\nu_1, \nu_2, \cdots, \nu_k} \left( t \right) = 1$). 
The values of the $c_{2k}$  are not yet established for ${}^{166} \textrm{Er}$. 
We therefore tacitly assume in the main text, when calculating concrete numerical 
examples for ${}^{166} \textrm{Er}$, that the above condition also still holds,  
 despite the prefactor $S^2$ enhancing the importance of spin-spin interactions in $\Phi_{0} \left( z, t \right)$. 
When this assumption is not applicable, one is required to take into account the time dependence of $\Phi_{0} \left( z, t \right)$ due to $\boldsymbol{M} \left( t \right)$ 
together with magnetostriction due to $\Phi_{dd} \left( z, t \right)$, which will change the system size $L_z$ as a function of $t$. This will in turn change the integration domain and quasi-1D density $n \left( z, t \right) = \left\vert \Psi \left( z, t \right) \right\vert^2$ in Eq.~\eqref{Phi-dd-Lambda-dd-connec}, and incur also a changed time dependence of $\Lambda'_{dd} \left( t \right)$, and the solution of the coupled system of equations \eqref{M-LLG-eq} and \eqref{GP_SW_nodiss-st2} needs to be found self-consistently. 
%}

%\colb{
For a harmonic trap, due to the resulting inhomogeneity of $\left\vert \Psi \left( z, t \right) \right\vert^2$, $\Phi_{dd} \left( z, t \right)$ will have more significant spatial dependence
than its box trap counterpart shown in Fig.~\ref{1D-Phidd}. Here, we note that Ref.~\cite{sapina_ground-state_2010} has already shown, for a spin-polarized gas, 
%,  $c_{2k}$ couplings not included),  
that magnetostriction occurs in a harmonic trap. 
The effect of magnetostriction is generally expected to be larger in a harmonic trap when  compared to a box trap with similar geometrical and dynamical parameters 
for large relative system size $L_z / l_{\perp}\gg 1$, 
at least under the above condition that the $S^2 c_{2k} / c_0$ %(for $1 \le k \le S$) 
are sufficiently small.
%}

\end{widetext}

\bibliography{LLG_SW_v45}

\end{document}